\documentclass[11pt,a4paper]{article} 
\usepackage{style}

\usepackage{amsmath,amssymb,epsfig,bm,amsfonts,yfonts,bbm,mathrsfs}

\title{Hydrodynamics and Jets in Dialogue}

\author{Stefan Floerchinger}
\emailAdd{stefan.floerchinger@cern.ch}
\author{and Korinna C. Zapp}
\emailAdd{korinna.christine.zapp@cern.ch}
\abstract{Energy and momentum loss of jets in heavy ion collisions can affect the fluid dynamic evolution of the medium. We determine realistic event-by-event averages and correlation functions of the local energy-momentum transfer from hard particles to the soft sector using the jet-quenching Monte-Carlo code \textsc{Jewel} combined with a hydrodynamic model for the background. The expectation values for source terms due to jets in a typical (minimum bias) event affect the fluid dynamic evolution mainly by their momentum transfer. This leads to a small increase in flow. The presence of hard jets in the event constitutes only a minor correction.}

\begin{document}
\maketitle

\newpage
\section{Introduction}

For a phenomenological, fluid dynamic description of heavy ion collisions it is usually assumed that the bulk of the medium produced after a heavy ion collision is in local thermal equilibrium. While this may be a reasonable approximation for the low-$\pt$ part, it is phenomenologically clear that high-$\pt$ particles do not originate from a locally equilibrated or thermal distribution. It is an interesting question how this non-equilibrated part influences the hydrodynamical evolution of the bulk. The most important effect in this regard might be due to the energy loss of high-$\pt$ particles propagating in the medium (jet quenching) which leads to a transfer of energy and momentum to the bulk part described by fluid dynamics. This could have important implications for the interpretation of soft observables (e.g. the anisotropic flow coefficients $v_n$) as well as jet measurements~\cite{Apolinario:2012cg}, which rely on background subtraction techniques assuming currently that the soft event and jets are uncorrelated. On the other side one expects that local fluid properties determine the strength of the energy loss. Generally speaking, energy loss is expected to be stronger for a denser medium. More specific, the jet quenching parameter $\hat q$ is expected to depend on the temperature and other parameters of the medium. 

The energy loss of hard partons due to induced gluon radiation in a hydrodynamic background has been studied in various approaches~\cite{Bass:2008rv,Chen:2011vt,Majumder:2011uk,Xu:2014ica}. They are, however, not suited for quantifying the energy and momentum deposition into the bulk. Firstly, they operate in a high energy limit, where there is no collisional energy loss, and secondly, they do not keep track of radiated gluons. It thus has to be assumed that all radiated energy gets dissipated locally, which is clearly a bad approximation for energetic emissions. This is different in Monte Carlo codes aiming at a consistent description of the entire jet and its interactions in a background described by hydrodynamics~\cite{Lokhtin:2008xi,Renk:2009nz,Schenke:2009gb}. They can trace all radiated partons, but also here the interactions between the jets and the bulk are accounted for in an effective way that cannot easily be translated into a local energy and momentum transfer between jets and background. 

The influence of jets on the hydrodynamic evolution of the bulk was first discussed in the context of Mach cone formation~\cite{CasalderreySolana:2004qm,Chaudhuri:2005vc,Neufeld:2008fi,Betz:2008js,Neufeld:2010tz,Betz:2010qh} and has been extended recently to other observables~\cite{Tachibana:2014lja,Andrade:2014swa}. Within AdS/CFT the interplay between the energy loss of a heavy quark and hydrodynamic excitations has been discussed in detail, see~\cite{CasalderreySolana:2011us} for an overview. In that context the source term can be extracted unambiguously from the energy and momentum loss of a (single) heavy quark, but the usual caveats concerning implications for QCD persist. The studies in QCD, on the other hand, feature an elaborate treatment of the hydrodynamic side of the problem, but have very simplified models for the energy and momentum deposited by the jets.

In general, a fully self-consistent description of the soft medium and high-$\pt$ part of the spectrum with its mutual interactions in QCD can be a rather difficult task (first steps in this direction are taken by transport codes~\cite{Li:2010ts,Ma:2010dv,Uphoff:2014cba}). On the other side, the phenomenological success of the current fluid dynamic model, which neglects the influence of non-equilibrated hard particles completely, suggests that the influence of the latter is not too large. In order to investigate this question more quantitatively, we employ here a non-self-consistent description where the bulk medium is first described in terms of conventional fluid dynamics neglecting non-thermal components. This leads in particular to a temperature and fluid velocity profile as a function of the space-time coordinates. In a second step we use these results to estimate the local transfer of energy and momentum from the hard particles to the medium. This results effectively in an additional source term in the fluid dynamic evolution equations. The influence of this source for fluid dynamics can then be estimated in a third step. The effect of a forth step, namely re-calculating the jets in the modified background, is expected to be numerically small and can thus be neglected.

In this study the jets are simulated with \textsc{Jewel}~\cite{Zapp:2012ak}, which employs a microscopic description for the interactions of the jets in the background. As a first approximation one can thus interpret the energy and momentum flow in the individual scattering processes as the energy-momentum exchange between the jets and the background. This provides a realistic and well constrained model for the local energy-momentum transfer to the bulk.

This paper is organized as follows. In section \ref{sec:TheHydrodynamicEvolution} we discuss the fluid dynamic evolution of the bulk including source terms for energy and momentum transferred from the hard sector. In section \ref{sec3} we introduce the description of jets and in the subsequent section \ref{sec:CharacterisingSourceTerm} we quantify the local energy and momentum transfer in terms of expectation values and correlation functions. Finally, we draw some conclusions in section \ref{sec:Conclusions}.
\section{The hydrodynamic evolution}
\label{sec:TheHydrodynamicEvolution}

In this section we discuss our formalism on the fluid dynamic side in more detail. We start from the fluid dynamic expression for the energy-momentum tensor
\begin{equation}
T^{\mu\nu}_\text{bulk} = (\epsilon+p+\pi_\text{bulk}) u^\mu u^\nu + (p+\pi_\text{bulk}) g^{\mu\nu} + \pi^{\mu\nu},
\label{eq:TmunuBulk}
\end{equation}
where $\epsilon$ is the energy density, $p$ is the pressure, $u^\mu$ is the fluid velocity, $\pi^{\mu\nu}$ is the shear stress tensor and $\pi_\text{bulk}$ is the bulk viscous pressure (we use signature $(-,+,+,+)$). Eq.\ \eqref{eq:TmunuBulk} accounts for the bulk contribution to the total energy-momentum tensor and gets supplemented by a similar contribution from the non-equilibrated, hard part of the medium $T^{\mu\nu}_\text{hard}$. The total energy-momentum tensor is conserved,
\begin{equation}
\partial_\mu (T^{\mu\nu}_\text{bulk} + T^{\mu\nu}_\text{hard}) = 0.
\end{equation}
We define the effective source terms for the bulk evolution by
\begin{equation}
J^\nu = - \partial_\mu T^{\mu\nu}_\text{hard}
\end{equation}
such that the energy-momentum conservation equation becomes
\begin{equation}
\partial_\mu T^{\mu\nu}_\text{bulk} = J^\nu.
\label{eq:EMConservation}
\end{equation}

When projected in the direction of the fluid velocity, this leads to the evolution equations for energy density
\begin{equation}
u^\mu \partial_\mu \epsilon + (\epsilon + p) \partial_\mu u^\mu - u_\nu \partial_\mu \pi^{\mu\nu} + \pi_\text{bulk} \partial_\mu u^\mu = - u_\nu J^\nu.
\label{eq:fluidEvo1}
\end{equation}
The third and forth term on the left hand side of eq.\ \eqref{eq:fluidEvo1} give the increase in thermal energy due to the shear and bulk viscous dissipative effects, respectively. This can be understood as a transfer of energy from the mechanical motion of the fluid to its thermal energy.
The second law of thermodynamics implies that this leads to an increase of entropy. In a first order fluid dynamic formalism, the constitutive relations
\begin{equation}
\begin{split}
\pi^{\mu\nu} = & - 2 \eta P^{\mu\nu}_{\;\;\;\alpha\beta} \partial^\alpha u^\beta,\\
\pi_\text{bulk} = & - \zeta \partial_\mu u^\mu
\end{split}
\label{eq:constitutiveRelations}
\end{equation}
(where $P^{\mu\nu}_{\;\;\;\alpha\beta} = \frac{1}{2}(\Delta^\mu_{\;\;\alpha} \Delta^\nu_{\;\;\beta} + \Delta^\mu_{\;\;\beta} \Delta^\nu_{\;\;\alpha}) - \frac{1}{3} \Delta^{\mu\nu} \Delta_{\alpha\beta}$ is the projector to the transverse and traceless part and $\Delta^{\mu\nu} = u^\mu u^\nu + g^{\mu\nu}$ is the projector orthogonal to the fluid velocity) make sure that this is indeed the case.

In a very similar way, the term on the right hand side of eq.\ \eqref{eq:fluidEvo1} accounts for a change of the thermodynamic internal energy (and therefore enthalpy and entropy) due to energy loss of high-momentum particles. This is a dissipative process, as well, and a consistent fluid dynamic description with the second law of thermodynamics requires
\begin{equation}
u_\nu J^\nu = g_{\mu\nu} u^\mu J^\nu \leq 0.
\label{eq:increaseEntropyRequirement}
\end{equation}
In the local fluid rest frame where $u^\mu = (1,0,0,0)$ this implies $J^0 \geq 0$, i. e. energy must be transferred from the non-equilibrated particles to the fluid (and not the other way). We will check below to what extend Eq.\ \eqref{eq:increaseEntropyRequirement} is indeed satisfied in currently used Monte-Carlo simulations of jet energy loss.

Let us now consider the remaining set of equations that follows from the conservation of energy-momentum, eq.\ \eqref{eq:EMConservation}. It is obtained by projecting to the direction orthogonal to the fluid velocity and yields an evolution equation for the latter,
\begin{equation}
(\epsilon + p + \pi_\text{bulk}) u^\mu \partial_\mu u^\alpha + \Delta^{\alpha \beta} \partial_\beta (p+\pi_\text{bulk}) + \Delta^\alpha_{\;\;\nu} \partial_\mu \pi^{\mu\nu} = \Delta^\alpha_{\;\;\nu} J^\nu.
\label{eq:fluidEvo2}
\end{equation}
In this equation, the second term on the left hand side accounts for an acceleration of the fluid due to pressure gradients. The third term accounts for the change in the fluid velocity due to dissipation of macroscopic kinetic energy into thermal energy. This leads usually to a damping of the fluid motion. The term on the right hand side of Eq.\ \eqref{eq:fluidEvo2} is a force term that accounts for the acceleration of the fluid due to high energetic particles that propagate through it. This is the force opposing drag.

The fluid dynamic evolution equations \eqref{eq:fluidEvo1} and \eqref{eq:fluidEvo2} have to be supplemented by constitutive equations for the shear stress tensor and bulk viscous pressure. In a first order (Navier-Stokes type) formalism these are of the form \eqref{eq:constitutiveRelations}, in a second order formalism these equations get supplemented by relaxation time terms. To solve the evolution equations one also needs an equation of state that relates pressure and energy density as well as the transport coefficients $\eta$ and $\zeta$ (and possible further coefficients such as relaxation times). 

\medskip

So far, we have not yet specified the source terms on the right hand side of eqns.\ \eqref{eq:fluidEvo1} and \eqref{eq:fluidEvo2}. If these correspond to high-momentum, non-equilibrated particles, they are in general different for each event. One might attempt at this point to implement an event-by-event description of fluid dynamics and a model for the high-momentum particles coupled to each other. On a technical level this becomes quickly rather involved. There is also a conceptual difficulty of drawing a line between the high-momentum part of the medium that is usually described in a microscopic way in terms of single particle excitations or partons and the low-momentum part that is described in a more macroscopic way in terms of fluid dynamics. 

We follow here another approach that uses a statistical description also for the non-equilibrated high-momentum part. More specific, we describe the influence of the non-equilibrated part of the medium onto the fluid dynamic variables and evolution in terms of a \emph{statistical ensemble} of sources $J^\nu$ or, equivalently, of a source component parallel to the fluid velocity,
\begin{equation}
J_S = u_\nu J^\nu
\end{equation}
and orthogonal to it,
\begin{equation}
J_V^\mu = \Delta^\mu_{\;\;\nu} J^\nu,
\end{equation}
as they appear on the right hand side of eqns.\ \eqref{eq:fluidEvo1} and \eqref{eq:fluidEvo2}. A particular configuration for a single event corresponds to one element of this ensemble. One possibility to characterize such an ensemble is in terms of a functional probability density
\begin{equation}
p[J_S, J_V]
\end{equation}
which is a functional of the source components $J_S(x)$ and $J_V(x)$ for a single event. Another possibility is in terms of the correlation functions or moments of this distribution, i. e.
\begin{equation}
\langle J_S(x) \rangle, \quad\quad \langle J_S(x) J_S(y) \rangle, \quad \quad \ldots
\end{equation}
and similar for $J_V$ and cross-terms. For fluctuations that are approximately Gaussian one can map the two characterizations  to each other; the properties of a Gaussian distribution are fixed uniquely in terms of its expectation values and two-point correlation functions. In that case our description amounts to splitting $J_S = u_\nu J^\nu$ and $J_V^\mu = \Delta^\mu_{\;\;\nu} J^\nu$ into expectation values
\begin{equation}
\langle J_S(x) \rangle, \quad\quad\quad \langle J_V^\mu(x) \rangle,
\label{eq:expvaluessource}
\end{equation}
and statistical Gaussian noise terms that are characterized in terms of the correlation functions
\begin{equation}
C_{SS}(x,y) = \langle J_S(x) J_S(y) \rangle, \quad C_{SV}^\mu(x,y) = \langle J_S(x) J_V^\mu(y) \rangle, \quad
C_{VV}^{\mu\nu}(x,y) = \langle J_V^\mu(x) J_V^\nu(y) \rangle.
\label{eq:corrfunctsource}
\end{equation}
It is clear that these objects depend on the details of the ensemble of events considered, for example collision energy, centrality and so on.

Let us now specialize our considerations to a situation with Bjorken boost and azimuthal rotation symmetry. For the fluid dynamic fields (enthalpy density $w=\epsilon+p$, fluid velocity $u^\mu$, shear stress $\pi^{\mu\nu}$ and bulk viscous pressure $\pi_\text{bulk}$) this implies that they can depend only on Bjorken time $\tau$ and radius $r$ (but not on rapidity $\eta$ and azimuthal angle $\phi$). For the fluid velocity only the components $u^\tau$ and $u^r$ can be non-zero and similar for the shear stress tensor $\pi^{\mu\nu}$. For the ensemble of sources $J_S$ and $J_V$ we assume that Bjorken boost and azimuthal rotation invariance are realized in a {\it statistical sense}.  For the expectation values in \eqref{eq:expvaluessource} this has the same implications as for the hydrodynamical fields. For the correlation functions in \eqref{eq:corrfunctsource} the situation is more complicated since they can depend also on the differences in rapidity and azimuthal angle between the two space-time points. 

We concentrate now on the fluid dynamic equations for an averaged situation where the source term on the right hand side of \eqref{eq:fluidEvo1} is replaced by the expectation value $\bar J_S=\langle J_S \rangle$. It reads
\begin{equation}
\begin{split}
& u^\tau \partial_\tau \epsilon + u^r \partial_r \epsilon + (\epsilon+p+\pi_\text{bulk}) (\partial_\tau u^\tau + \partial_r u^r + \tfrac{1}{\tau} u^\tau + \tfrac{1}{r} u^r)\\
& + u^\tau \left[ \partial_\tau \pi^{\tau\tau} + \tfrac{1}{\tau} \pi^{\tau\tau} + \partial_r \pi^{\tau r} + \tfrac{1}{r} \pi^{\tau r} + \tfrac{1}{\tau} \pi^{\eta\eta} \right]\\
& - u^r \left[ \partial_\tau \pi^{\tau r} + \tfrac{1}{\tau} \pi^{\tau r} + \partial_r \pi^{rr} + \tfrac{1}{r} \pi^{rr} - \tfrac{1}{r} \pi^{\phi\phi}\right] = - \bar J_S.
\end{split}
\label{eq:2.15}
\end{equation}
Similarly, the right hand side of Eq. \eqref{eq:fluidEvo2} is replaced by $\bar J_V^\alpha = \langle J_V^\alpha \rangle$. The components $u^\phi$ and $u^\eta$ vanish due to symmetries and as a result of the constraint $u_\mu u^\mu = -1$ only one of the remaining equations (say for $u^r$) is independent. It reads
\begin{equation}
\begin{split}
& (\epsilon + p + \pi_\text{bulk}) (u^\tau \partial_\tau u^r + u^r \partial_r u^r) + u^r u^\tau \partial_\tau (p+\pi_\text{bulk}) + (u^\tau)^2 \partial_r (p+\pi_\text{bulk}) \\
& - u^\tau u^r \left[ \partial_\tau \pi^{\tau\tau} + \tfrac{1}{\tau} \pi^{\tau\tau} + \partial_r \pi^{r\tau} + \tfrac{1}{r} \pi^{r\tau} + \tfrac{1}{\tau} \pi^{\eta\eta} \right] \\
& + (u^\tau)^2 \left[ \partial_\tau \pi^{\tau r} + \tfrac{1}{\tau} \pi^{\tau r} + \partial_r \pi^{rr} + \tfrac{1}{r} \pi^{rr} - \tfrac{1}{r} \pi^{\phi\phi}\right] = \bar J_V^r.
\end{split}
\label{eq:2.16}
\end{equation}

Eqns.\ \eqref{eq:2.15} and \eqref{eq:2.16} determine the time evolution of energy density $\epsilon(\tau, r)$ and (the radial component of) the fluid velocity, respectively. They depend on the shear stress components $\pi^{\mu\nu}$ and the bulk viscous pressure. We neglect here the latter while the former is determined from a time evolution equation that supplements the first line of eq. \eqref{eq:constitutiveRelations} with a relaxation time term $\sim \tau_\text{Shear}$. The shear viscosity $\eta$ and the relaxation time $\tau_\text{Shear}$ are chosen for concreteness according to their AdS/CFT values $\eta = s/(4\pi)$, $\tau_\text{Shear} = (2-\ln 2)/(2\pi T)$. Somewhat larger values as they are presumably more realistic for QCD would not change our findings substantially. To solve eqns.\ \eqref{eq:2.15} and \eqref{eq:2.16} one also needs a thermodynamic equation of state that relates pressure and energy density to the temperature. We take the parametrization s95p-PCE  of ref.\ \cite{Shen:2010uy}. For the initial conditions at time $\tau=\tau_0=0.6 \, \text{fm}$ we follow ref.\ \cite{Qiu:2011hf} in assuming that the radial fluid velocity vanishes, $u^r=0$, that the shear stress assumes its Navier-Stokes value and that the transverse energy density is determined by the nuclear overlap function  for central collisions with maximal temperature $T=485\, \text{MeV}$.

\begin{figure}[ht]
\centering
\includegraphics[width=0.58\linewidth]{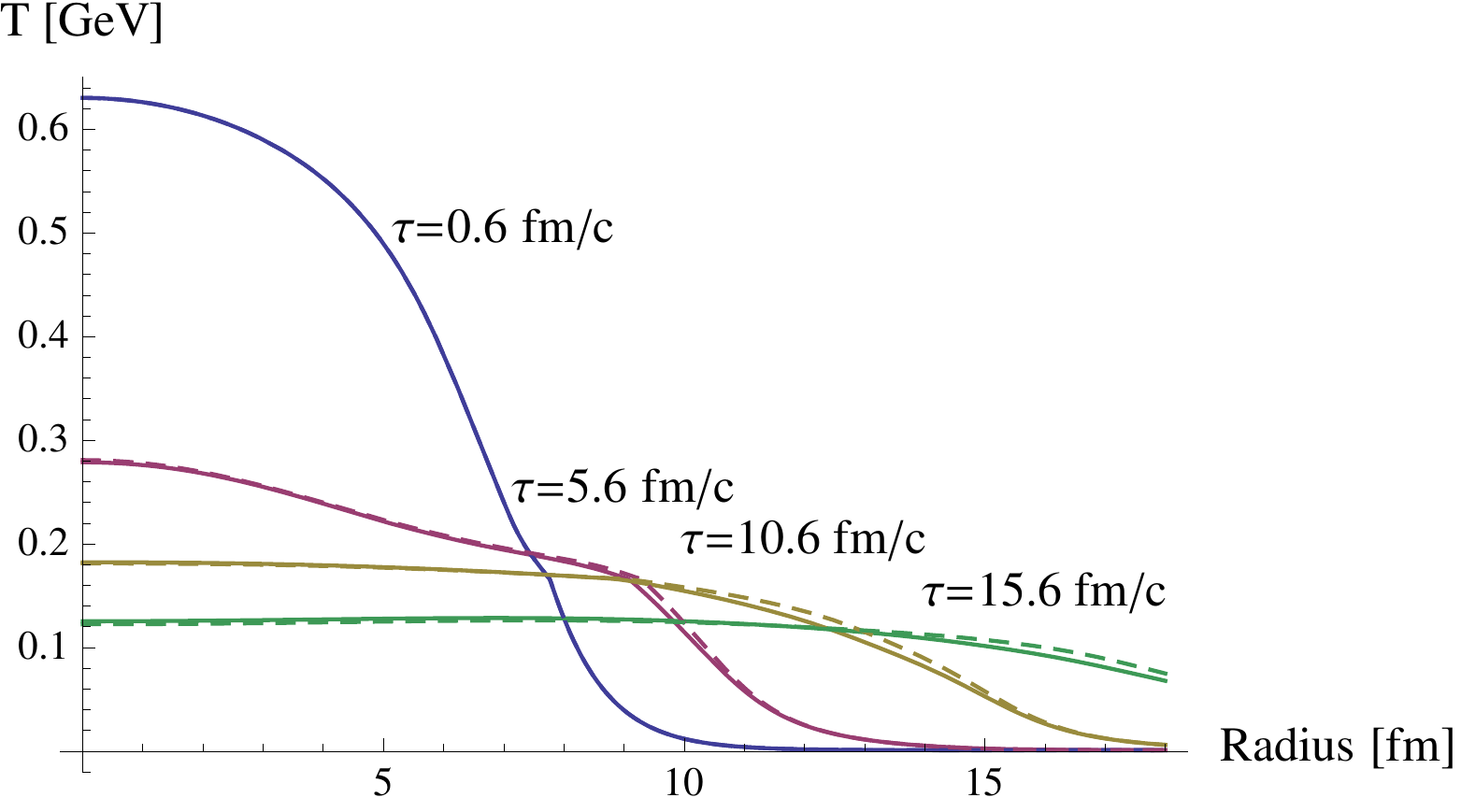}
\caption{Temperature as a function of radius for different times $\tau$. The solid lines correspond to vanishing source terms, $\bar J_S = \bar J_V = 0$ while the dashed lines correspond to the full result where they are taken into account as calculated in section \ref{sec:CharacterisingSourceTerm}.}
\label{fig::temp}
\end{figure}

\begin{figure}[ht]
\centering
\includegraphics[width=0.58\linewidth]{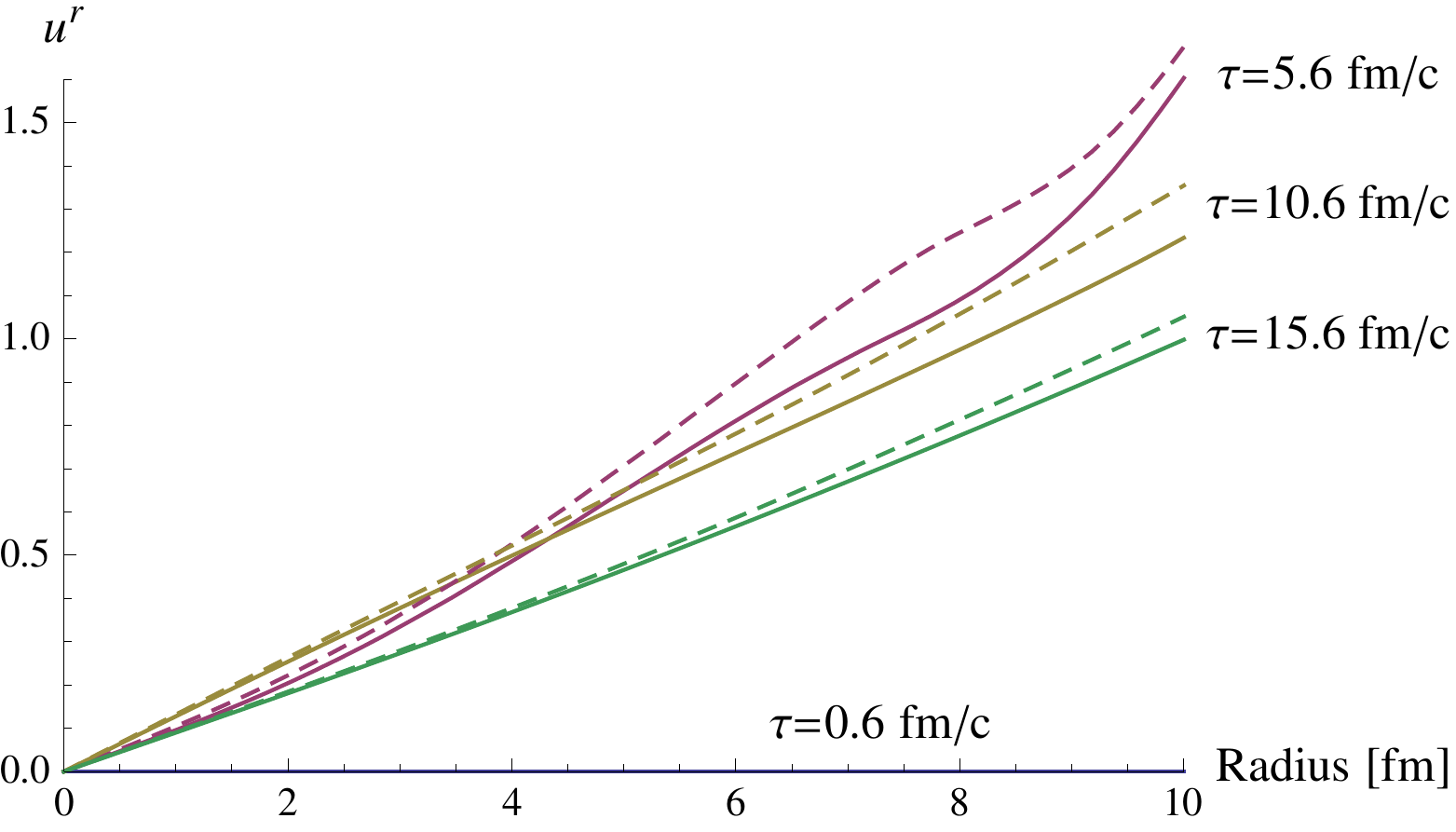}
\caption{Radial component of the fluid velocity $u^r$ as a function of radius for different times $\tau$. The solid lines correspond to vanishing source terms, $\bar J_S = \bar J_V = 0$ while the dashed lines correspond to the full result where they are taken into account as calculated in section \ref{sec:CharacterisingSourceTerm}.}
\label{fig::fluidVelocity}
\end{figure}

Once the differential eqns.\ \eqref{eq:2.15} and \eqref{eq:2.16} are solved, one can use the equation of state to extract the temperature. The result is shown as a function of radius for different times $\tau$ in Fig.\ \ref{fig::temp}. The solid lines give the result without source terms, i.\ e.\ for $\bar J_S = \bar J_V = 0$, while the dashed lines correspond to the full result with expectation values for sources for a 'typical' event calculated as described in sect.\ \ref{sec:CharacterisingSourceTerm}.
In fig.\ \ref{fig::fluidVelocity} we plot the radial fluid velocity as a function of radius in a similar way.
One observes that the source terms $\bar J_S$ and $\bar J_V$ have two effects: One is a slight increase in temperature at small radii and at early times, which is an expected effect from dissipation. The other is an increase of radial flow at intermediate times, the jets drag the fluid outwards. The slight decrease of the temperature in the centre and the increase at large radii at later times are also a consequence of the larger flow. Both effects are relatively small. The change in the (averaged) temperature evolution seems to be negligible for practical purposes. The effect of the additional dissipated energy is hardly visible in fig.\ \ref{fig::temp} while the larger radial flow leads at larger radii to an increase in temperature of a few percent. For the radial component of the fluid velocity this effect is more direct and leads to an increase up to about $10\%$.

In our setup the expectation values $\bar J_S$ and $\bar J_V$ are by construction symmetric under azimuthal rotations and can therefore not contribute to the harmonic flow coefficients $v_m$. The effect of energy and momentum transfer from jets to the medium on these observables is encoded in correlation functions as in eq.\ \eqref{eq:corrfunctsource}. We plan to investigate this more quantitatively in a separate publication.
\section{Jet quenching in a hydrodynamic background}
\label{sec3}

Let us now describe the formalism we use for the description of jets. Jets are simulated in \textsc{Jewel}~\cite{Zapp:2012ak} with the hydrodynamic calculation presented in section~\ref{sec:TheHydrodynamicEvolution} as background. 

Geometrical aspects of the nucleus-nucleus collision are modeled using a Glauber model~\cite{Eskola:1988yh} with a Woods-Saxon nuclear potential. Once the impact parameter $b$ chosen according to the geometrical cross section is fixed, the nuclear overlap can be computed
\begin{equation}
T_{AB}(b) = \int \d^2 \mathbf{r}\, \d z_1\, \d z_2\, n_A(\mathbf{r},z_1) n_B(\mathbf{r}-\mathbf{b},z_2) \,.
\end{equation}
Here, $n_A$ denotes the nuclear potential, $z$ is chosen along the beam axis and $\mathbf{r}$ and the impact parameter $\mathbf{b}$ are orthogonal to this direction. The mean number of di-jets in the event is then given by
\begin{equation}
\mean{N_\text{di-jet}} = \sigma_\text{di-jet}(p_{\perp, \text{cut}}) T_{AB}(b) \,,
\end{equation}
where $\sigma_\text{di-jet}(p_{\perp, \text{cut}})$ is the cross section per nucleon-nucleon collision for the production of a di-jet with $p_{\perp, \text{jet}} > p_{\perp, \text{cut}}$. In \textsc{Jewel} the jet production matrix elements and initial state parton showers are simulated by \textsc{Pythia}\,6.4~\cite{Sjostrand:2006za}, which provides the leading order di-jet cross section with the EPS09 nuclear PDF set~\cite{Eskola:2009uj}. The  number of di-jets per event is Poisson distributed. The jets are produced at $z=0$ due to the strong Lorentz contraction of the colliding nuclei along the beam direction. In the transverse plane they are distributed according to the density of binary nucleon-nucleon collisions. In this set-up jets from different nucleon-nucleon interactions are uncorrelated.

The QCD evolution of the jets and re-scattering in the background take place on comparable time scales and are described in a common framework. It is assumed that the interactions of a jet resolve quasi-free partons in the medium and an infra-red continued version of the perturbative matrix elements can be used to describe them. \textsc{Jewel} uses leading order (LO) $2\to 2$ matrix elements to describe the re-scattering of hard partons in the medium and generates radiative corrections with the parton shower. The cross section for the re-scattering of a hard parton of type $i$ with energy $E$ in a background of temperature $T$ and fluid velocity $u^\mu$ is then given by
\begin{equation}
\label{eq::scatxec}
\sigma_i(E,T,u^\mu) = 
\int\limits_0^{|\hat t|_{\text{max}}(E,T,u^\mu)}\!\!\!\!\! \d |\hat t|\!\! 
 \int\limits_{x_{\text{min}}(|\hat t|)}^{x_{\text{max}}(|\hat t|)} \!\!\!\!\! \d x
\sum_{j \in \{\text{q,\=q,g}\}} \!
f_j^i(x, \hat t) \frac{\d \hat \sigma_j}{\d \hat t}(x\hat s,|\hat t|) \,,
\end{equation}
where the partonic PDFs $f_j^i(x,Q^2)$ encode possible initial state radiation off the energetic parton\footnote{In principle also the thermal scattering centre can emit such radiation, but this is neglected in the current \textsc{Jewel} implementation due to very limited phase space.}. Keeping the leading terms only and introducing the infra-red regulator $\mu_\text{D} \approx 3 T$ the partonic cross section reduces to

\begin{equation}
 \frac{\d \hat \sigma_j}{\d \hat t}(\hat s,|\hat t|) = C_j
\frac{\pi}{\hat s^2} \alphas^2(|\hat t| + \mu_{\text{D}}^2)\frac{\hat s^2 + (\hat
s-|\hat t|)^2}{(|\hat t| + \mu_{\text{D}}^2)^2} \longrightarrow C_j
2 \pi \alphas^2(|\hat t| + \mu_{\text{D}}^2)\frac{1}{(|\hat t| +
\mu_{\text{D}}^2)^2} \,,
\end{equation}
where $C_j$ is the appropriate colour factor. By adding the parton shower a systematic approximation to higher order $2\to n$ matrix elements is constructed. In this way both elastic and inelastic scattering processes are generated with the (leading log) correct relative rates.

The parton shower thus generates all emissions -- those associated to the QCD evolution of the jet (which would also take place in the absence of the background) and those initiated by re-scattering. 
In fact, it is generally  impossible to assign an emission to a particular scattering process. 
The interplay of competing sources of radiation as well as the LPM interference are governed by the formation times of the emissions. 
When two emissions take place at the time the one with the shorter formation is formed as an independent particle and all scattering process within the formation time of an emission act coherently. 

\smallskip

The local scattering rate is given by the product of the parton density and the scattering cross section~(\ref{eq::scatxec}) (taking care of the color factors for different parton species). When a scattering takes place a scattering centre is generated from the local thermal distribution and the scattering process is simulated explicitly. The scattering centers are dynamical and recoil against the hard parton. This allows to keep track of the energy and momentum exchange between the jet and the background.

\smallskip

When \textsc{Jewel} runs with the hydrodynamic background described in section~\ref{sec:TheHydrodynamicEvolution} it takes the temperature $T(x)$ and transverse fluid rapidity $\beta(x)$ (related to the radial component of the fluid velocity by $u^r=\sinh \beta(x)$) as input. The parton densities and momentum distributions are then computed assuming an ideal gas equation-of-state. 

\begin{figure}[ht]
\includegraphics[width=0.48\linewidth]{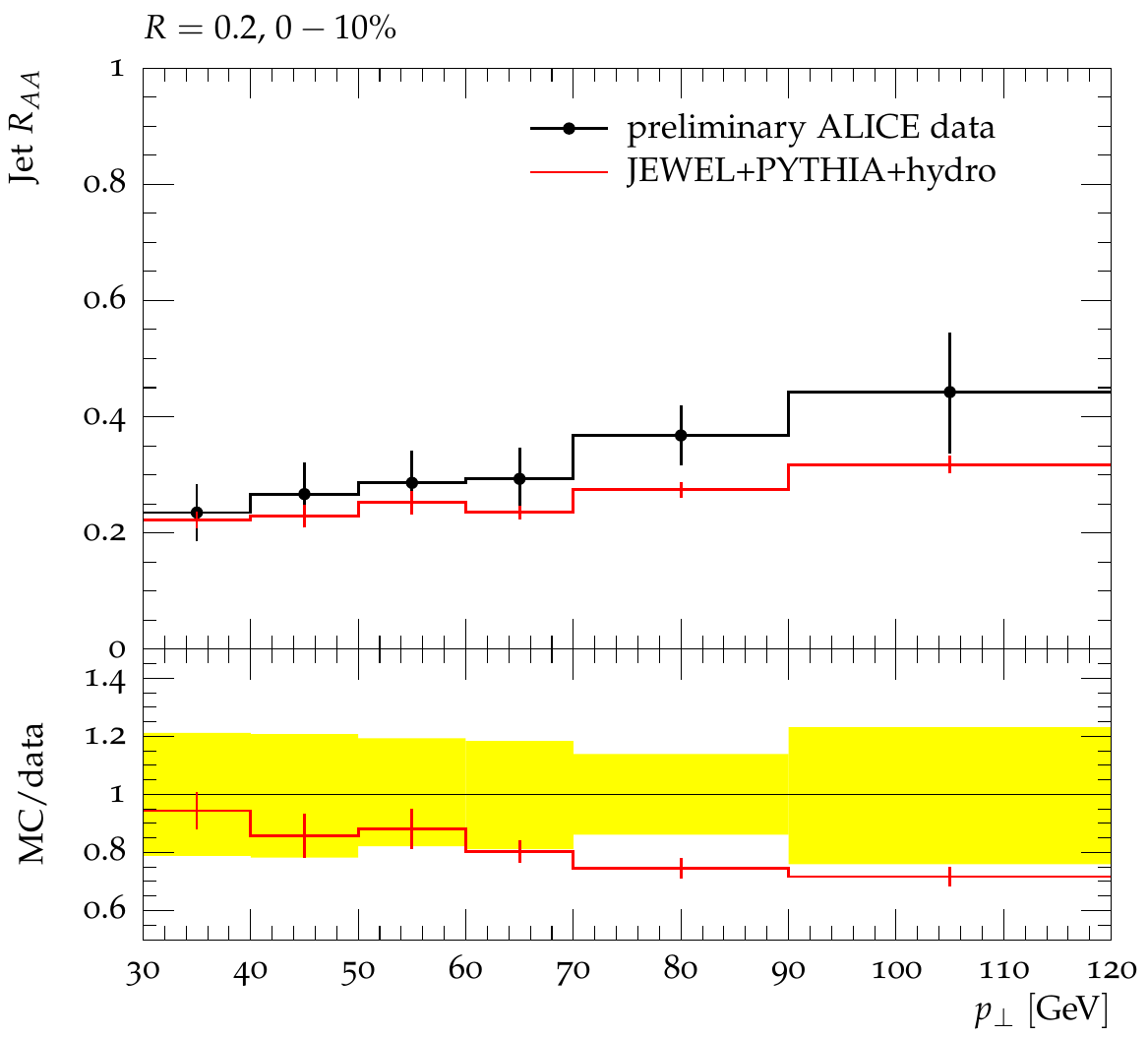}
\includegraphics[width=0.48\linewidth]{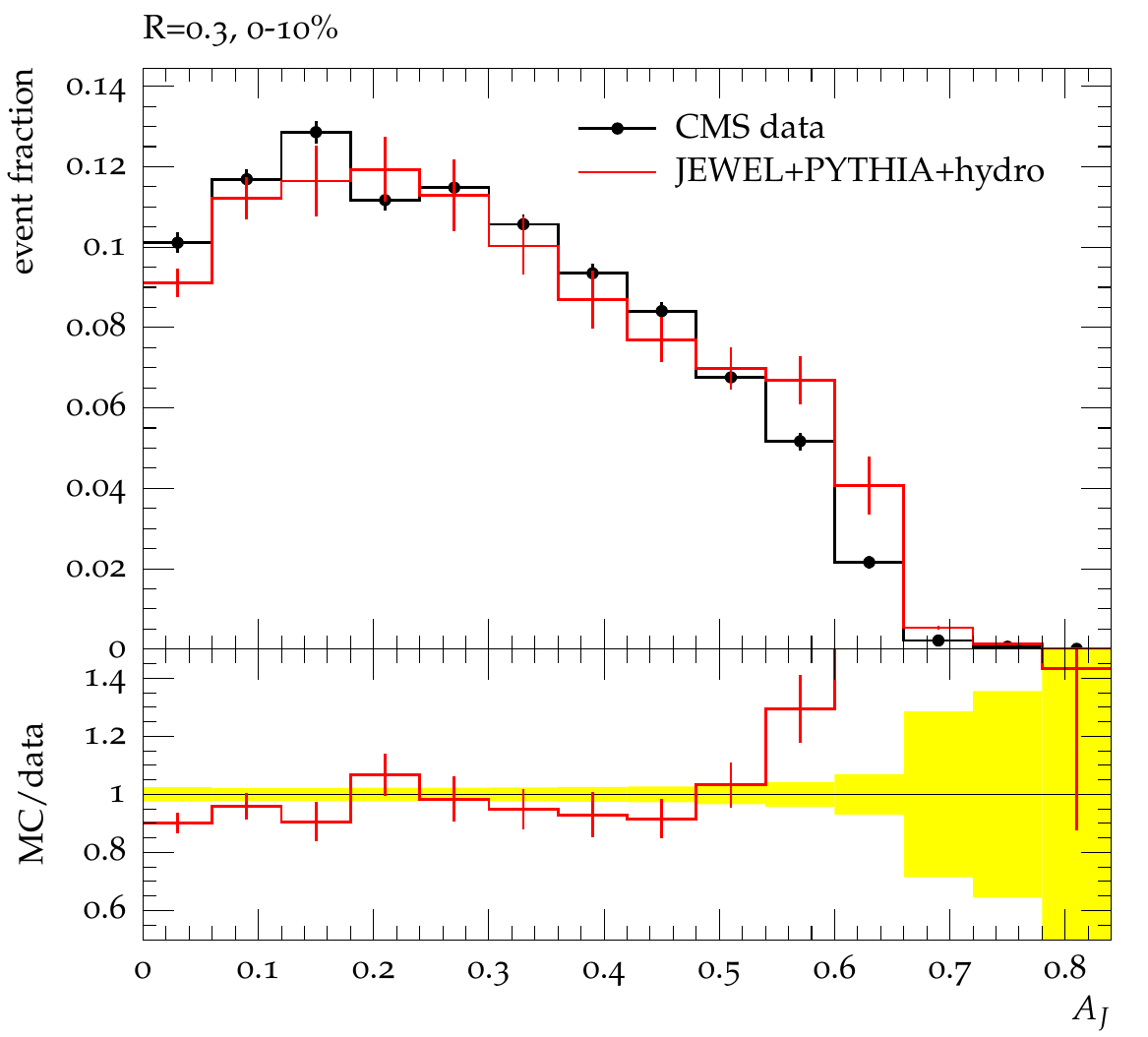}
\caption{\textbf{Left:} Nuclear modification factor of jets in Pb+Pb collisions at $\sqrt{s_\text{NN}} = \unit[2.76]{TeV}$ compared to preliminary  \textsc{Alice} data~\cite{Reed:2013rpa} (data points read off the plot, only maximum of statistical and systematic errors shown). Jets are reconstructed in $|\eta| < 0.5$ and are required to have a leading track with $\pt > \unit[5]{GeV}$.
\textbf{Right:} Di-jet asymmetry $A_J = (p_{\perp,1} - p_{\perp,2})/(p_{\perp,1} + p_{\perp,2})$ in Pb+Pb collisions at $\sqrt{s_\text{NN}} = \unit[2.76]{TeV}$ for transverse momenta of the leading jet $p_{\perp,1} > \unit[120]{GeV}$ . The sub-leading jet is required to have $p_{\perp,2} > \unit[30]{GeV}$ and $\df > 2π/3$. The \textsc{Cms} data~\cite{Chatrchyan:2012nia} are not unfolded for jet energy resolution, so the Monte Carlo events were smeared with the parametrisation from~\cite{Chatrchyan:2012gt}. In both plots the Monte Carlo results are for \unit[0]{\%} centrality. Monte Carlo events are analysed with \textsc{Rivet}~\cite{Buckley:2010ar} using \textsc{FastJet}~\cite{Cacciari:2011ma}.
}
\label{fig::validation}
\end{figure}

With the same parameter settings as used in~\cite{Zapp:2013vla} with the simple Bjorken-type background a very reasonable agreement with the jet quenching data is found. Figure~\ref{fig::validation} shows as examples the nuclear modification factor of jets ($R_\text{AA} = (\d \sigma^\text{(AA)}_\text{jet}/\d \pt)/(N_\text{coll}\cdot \d \sigma^\text{(pp)}_\text{jet}/\d \pt)$) and the di-jet asymmetry ($A_J = (p_{\perp,1} - p_{\perp,2})/(p_{\perp,1} + p_{\perp,2})$).

\section{Characterising the source term}
\label{sec:CharacterisingSourceTerm}

The four-momentum exchange between the jet and the background can be regarded as source term in the hydrodynamic evolution. In \textsc{Jewel} an event contains a discrete set of scatterings with momentum transfers $\Delta p^\mu_i$. The source term can thus be written as
\begin{equation}
J^\mu(x) = \sum_{i} \Delta p^\mu_i \delta^{(4)}(x-x_i) \,.
\end{equation}

We are now coming back to condition~(\ref{eq:increaseEntropyRequirement}), namely that the energy transfer has to be positive (i.e.\ from the jets to the background) in order not to violate the second law of thermodynamics. The phenomenologically relevant cases concern the propagation of a jet, which is by construction harder than the thermal background and will thus on average lose energy and momentum so that the source term is positive. In principle, the framework also allows to consider the propagation of very soft (compared to the thermal momentum scale) partons in the background. In this case the parton would on average gain energy through elastic scattering in the background (soft partons being unable to emit resolvable radiation) and the energy transfer becomes negative. This does not automatically violate the entropy condition~(\ref{eq:increaseEntropyRequirement}), as the contribution of the jet or soft parton to the total entropy has to be taken into account as well. This is non-trivial since these partons are out of equilibrium, but it is a generic expectation that in both cases (hard or soft partons) their (out-of-equilibrium) entropy increases.

\bigskip

For the most central collisions ($b=0$) the averaged source term $\mean{J^\mu}$ is azimuthally symmetric. It is, however, not boost invariant, since the jet production cross section is rapidity dependent\footnote{The jet production cross section depends on momentum rapidity, which is correlated to the space-time rapidity, since jets are produced at $t=0$ and at (or close to) $z=0$.} and the energy loss itself can in general also be rapidity dependent. For simplicity, we extract the source term only in the central unit of rapidity, where it varies only mildly, to preserve the symmetry of the background (the extension to a non-trivial rapidity dependence is straightforward). Consequently, $\mean{J^\mu}$ depends only on $\tau$ and $r$ and not on $\phi$ and $\eta$. The projections of $\mean{J^\mu}$ parallel and orthogonal to the fluid velocity are computed using for $u^\mu$ the solution to the hydrodynamic equations without the source term. This is a good approximation as long as the source term is small, when it is not small the procedure may be iterated using the new solution. Concerning the source term from jet energy loss, we assume the nucleon-nucleon collisions in a nucleus-nucleus event to be independent. Then the expectation values $\bar J_S$ and $\bar J_V^\mu$ scale trivially with the number of di-jets. The results presented in this section are averaged over $\phi$ and the central \mbox{unit in $\eta$}.

\begin{figure}[ht]
\centering
\includegraphics[angle=-90,width=0.65\linewidth]{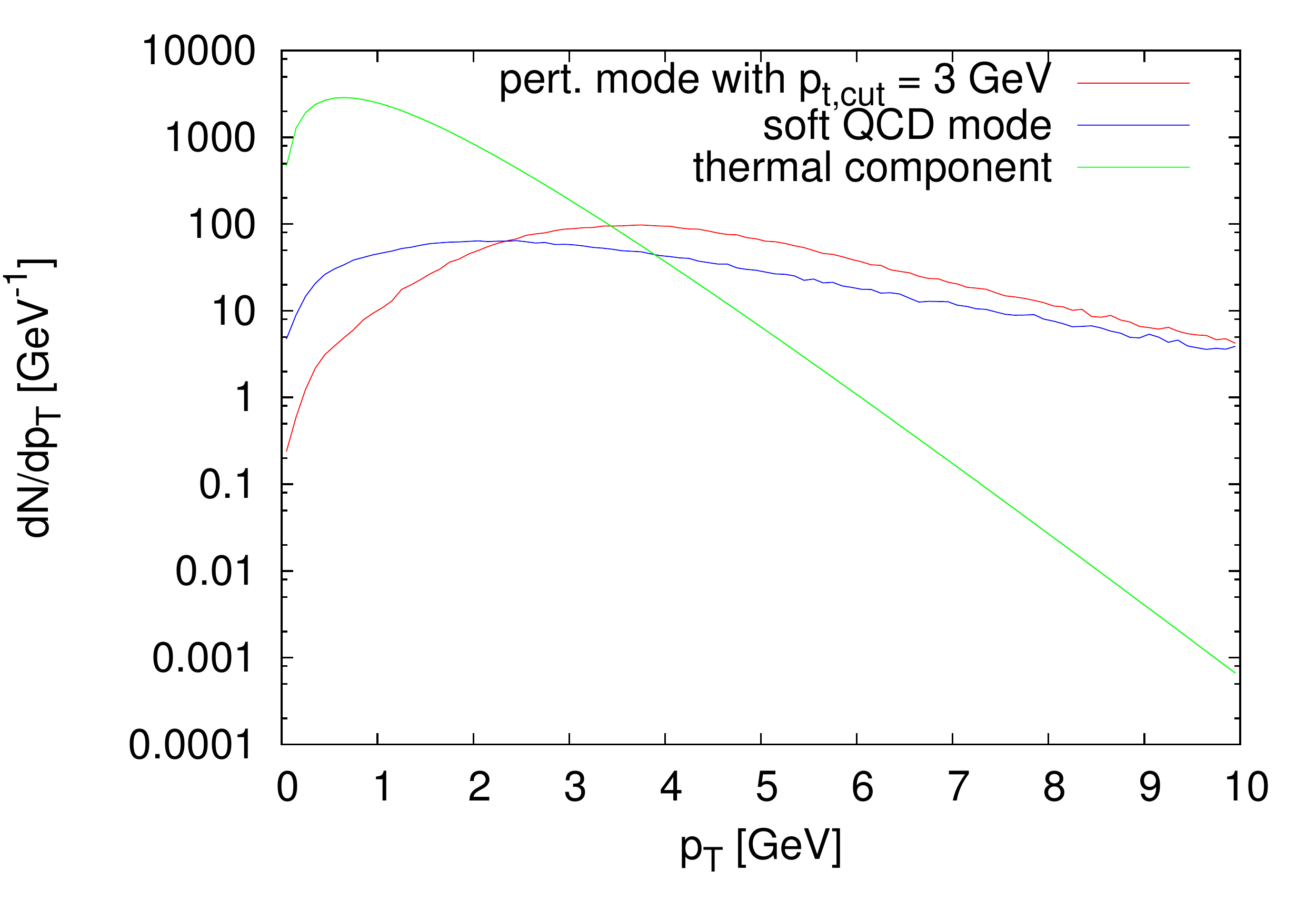}
\caption{Distribution of jets entering the final state parton evolution in the perturbative and soft QCD mode compared to the thermal parton population with $T=\unit[485]{MeV}$ in the central unit of rapidity.}
\label{fig::jetspec}
\end{figure}

We study Pb+Pb collisions at $\sqrt{s_\text{NN}}=\unit[2.76]{GeV}$ in two scenarios, namely (i) minimum bias collisions representing a typical event and (ii) events containing a $\mathcal{O}(\unit[100]{GeV})$ jet representing events in which a hard jet was triggered. In the former a major difficulty consists in defining the jet population. The perturbative jet cross section is infra-red divergent and has to be regularised, e.g.\ by a $\pt$ cut-off. Very low $\pt$ 'jets' should not be included anyway, because particles with momenta close to the thermal momentum at early times should be considered part of the background. The jet population should thus contain all jets that are harder than the thermal background. The $\pt$-cut has to be placed on the jet production matrix element. In the distribution of jets entering the final state evolution it appears smeared out due to momentum re-distribution in the parton shower. With a $\pt$-cut of $p_{\perp,\text{cut}} = \unit[3]{GeV}$ the initial jet distribution crosses the thermal distribution at its maximum, so that it populates predominantly the $\pt$ region where the jet population is larger than the thermal one (figure~\ref{fig::jetspec}). With this value of the cut-off the leading order di-jet cross section is $\sigma_\text{di-jet} = \unit[52.5]{mb}$, which nearly saturates the inelastic proton-proton cross section of $\sigma_\text{inel} = \unit[(62.8_{-4.0}^{+2.4}\pm 1.2)]{mb}$~\cite{Abelev:2012sea}. It can thus be expected that in this regime some mechanism sets in that unitarises the cross section (e.g.\ multi parton interactions). This is not quantitatively under control and we do not attempt not model it here, but it adds to the already large uncertainty of the leading order di-jet cross section. In \textsc{Pythia}\,6.4 a soft QCD mode is available, which regularises the jet cross section in a different way~\cite{Sjostrand:1987su} and also leads to a lower di-jet cross section of \unit[41.3]{mb}. This scenario is studied as an alternative to our default set-up to get an estimate of the uncertainties. The initial jet distribution (before final state parton showering) in the perturbative (default) and soft QCD scenario are compared to the thermal parton distribution in figure~\ref{fig::jetspec}.

\begin{figure}[ht]
\centering
\includegraphics[angle=-90,width=0.65\linewidth]{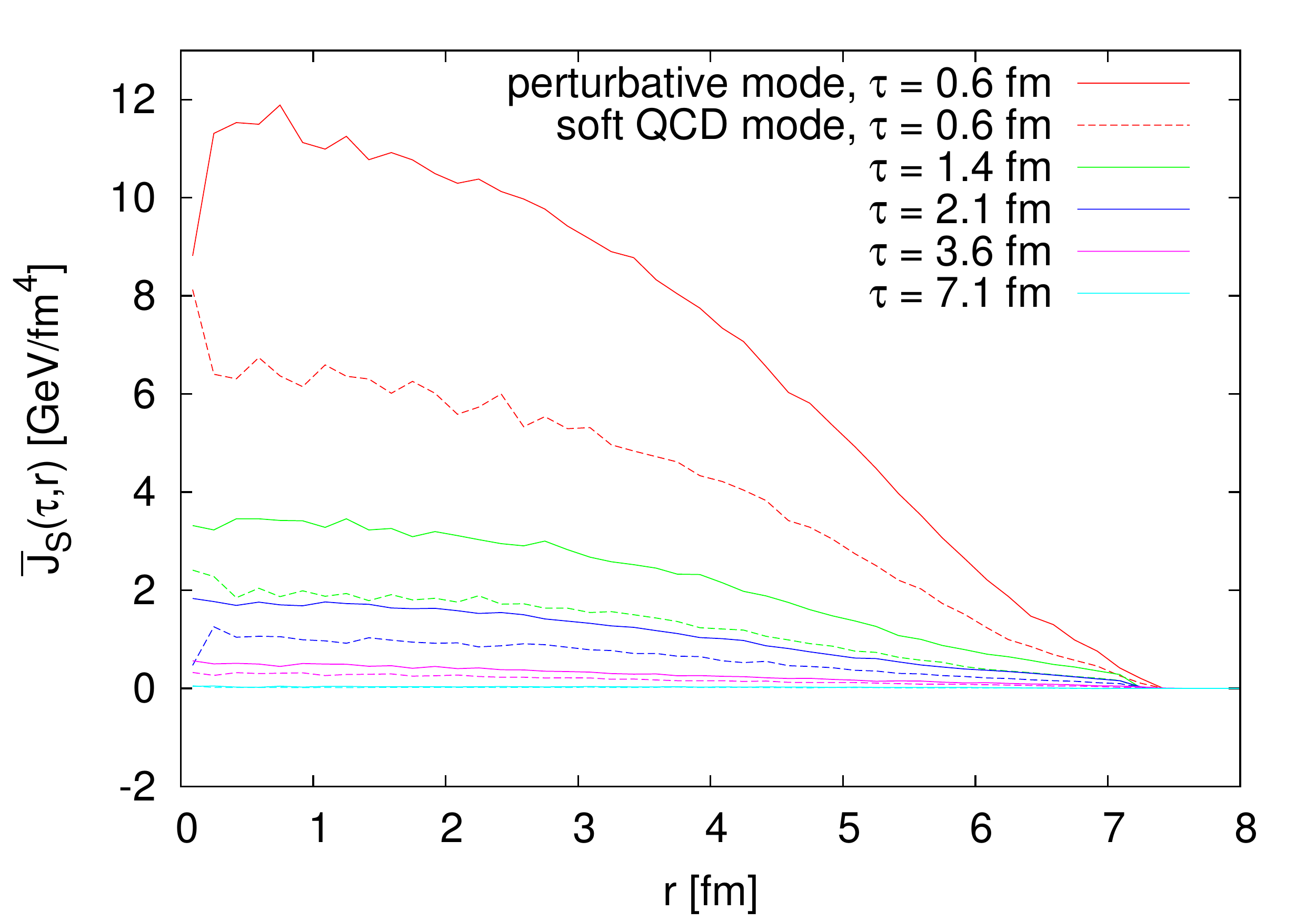}
\caption{Event-averaged source term $\bar J_S$ parallel to the fluid velocity in the perturbative and soft QCD mode for different values of $\tau$. $\bar J_S$ is averaged over the azimuthal angle $\phi$ and the central unit in rapidity $\eta$. }
\label{fig::2js}
\end{figure}

\begin{figure}[ht]
\includegraphics[angle=-90,width=0.49\linewidth]{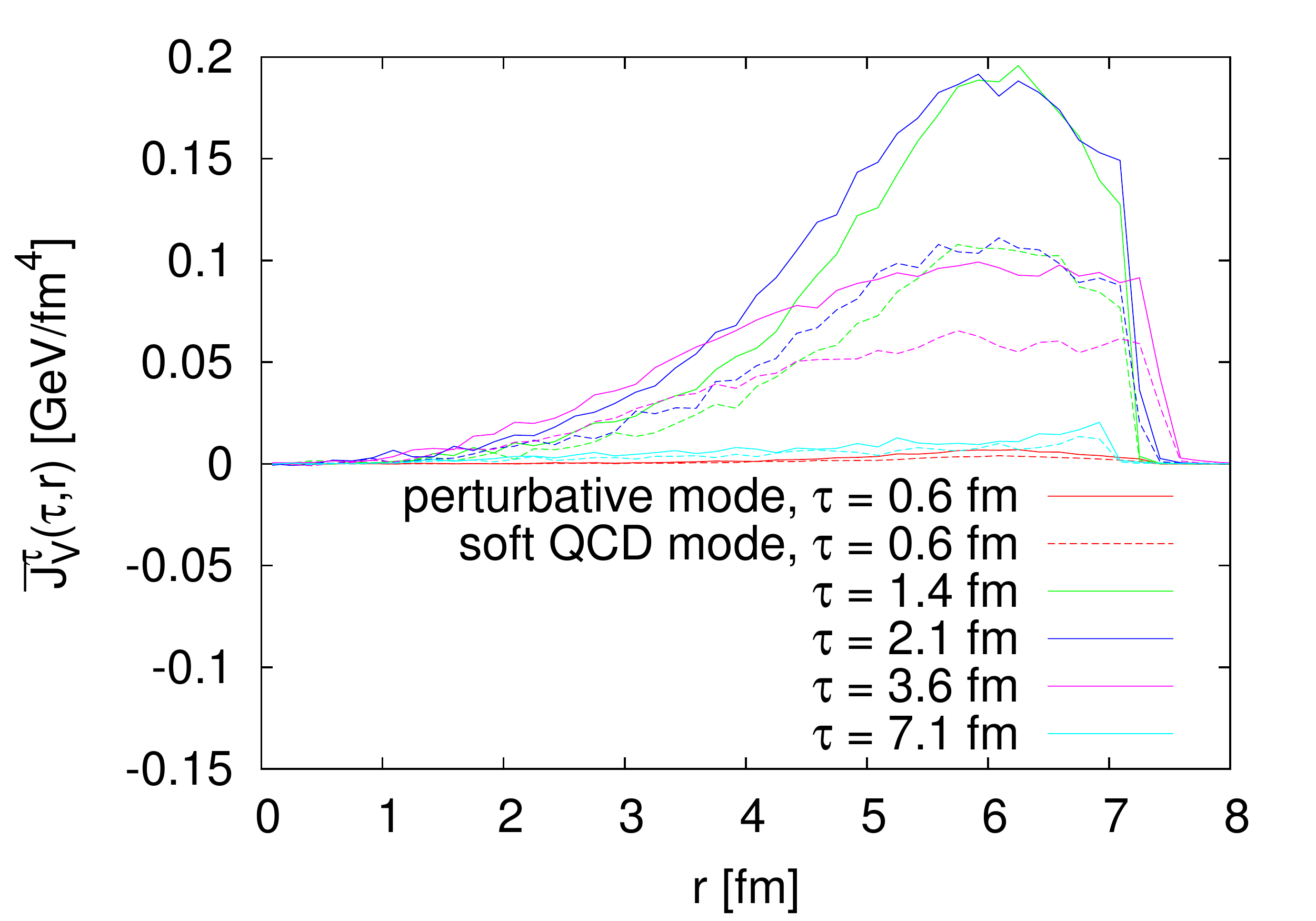}
\includegraphics[angle=-90,width=0.49\linewidth]{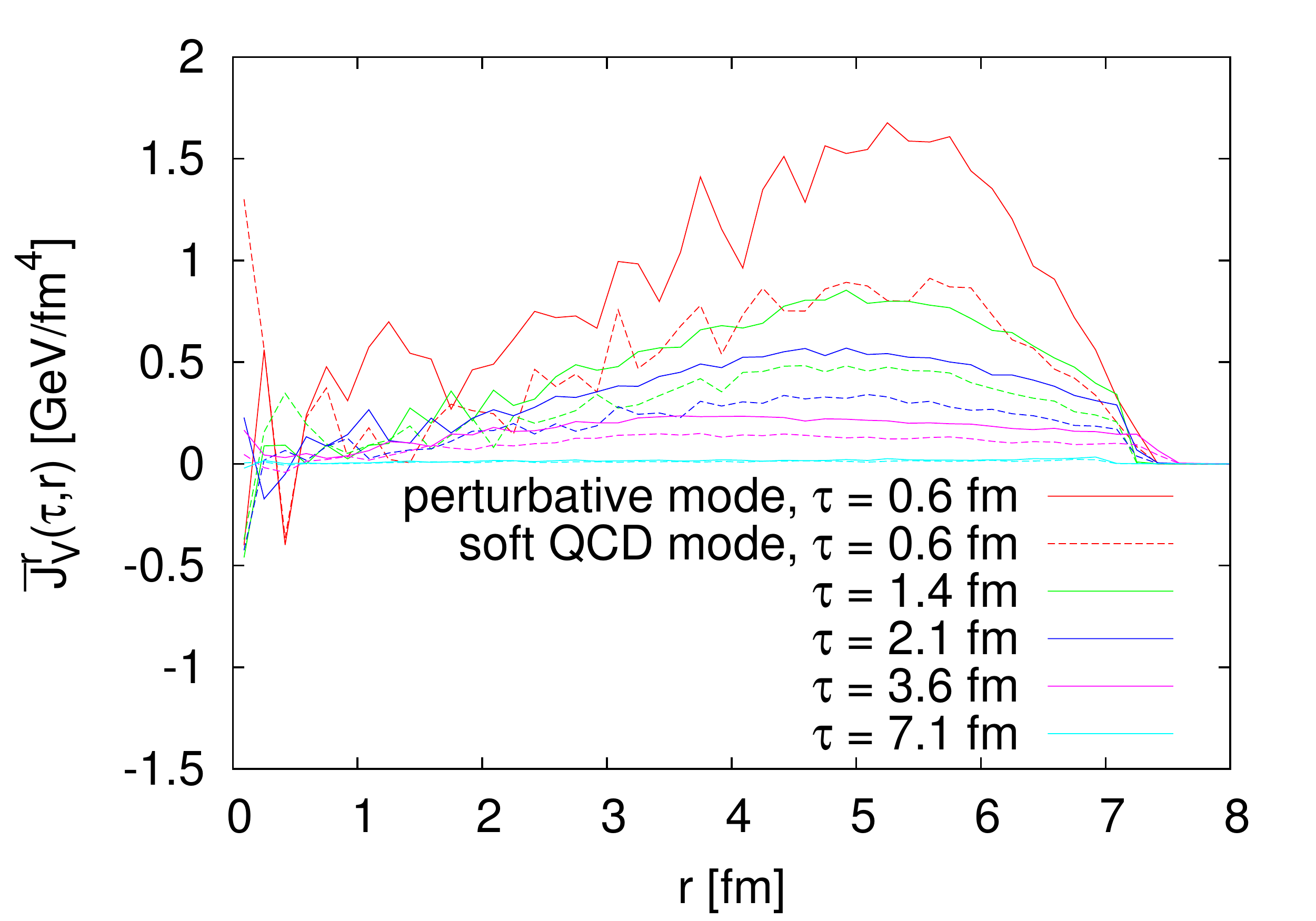}
\caption{Components $\bar J_V^\tau$ (left) and $\bar J_V^r$ (right) of the event-averaged source term orthogonal to the fluid velocity in the perturbative and soft QCD mode for different values of $\tau$. $\bar J_V^\tau$ and $\bar J_V^r$ are averaged over the azimuthal angle $\phi$ and the central unit in rapidity $\eta$.}
\label{fig::jv}
\end{figure}

In the soft QCD mode the average jet $\pt$ is smaller than in the perturbative mode. The energy and momentum deposited in the medium per jet is thus lower in the former. In addition, the number of jets per event is also smaller due to the smaller cross section. Figure~\ref{fig::2js} shows the projection $\bar J_S$ of the averaged source term on the fluid velocity per event. The results per event in the two cases differ by a factor $\sim 1.8$, but the shapes are very similar. The number of di-jets per event in the perturbative and soft QCD mode differs only by \unit[30]{\%} (1705 and 1342, respectively). The remaining difference between the source terms is due to the different $\mean{\pt}$. This also applies to the components orthogonal to the fluid velocity $\bar J_V^\tau$ and $\bar J_V^r$ shown in figure~\ref{fig::jv} ($\bar J_V^\phi$ and $\bar J_V^\eta$ vanish). The two components are related through the condition $u_\mu \bar J_V^\mu=0$. 

Initially, $\bar J_S$ scales roughly as $T^2(\tau,r)\cdot N_\text{coll}(r)$. This can be understood as follows: In the absence of strong flow (which is the case at mid-rapidity) the average energy loss per scattering is proportional to $T$. In \textsc{Jewel}, the scattering cross section in also temperature dependent since the infra-red regulator depends on temperature ($\mu_\text{D}\approx 3 T$). Together with the density of scattering centers, $n \sim T^3$, this leads to a linear temperature dependence of the scattering rate per hard parton. The factor $N_\text{coll}(r)$ comes from the initial distribution of jets. In practice, the source term $\bar J_S$ falls off faster with $\tau$ than $T^2(\tau,r)$: The average jet $\pt$ is only a few \unit{GeV} so that the jet partons evolve through splitting and scattering quickly to nearly thermal momenta and don't contribute to the source terms any more. In contrast to this $\bar J_V$ builds up at later times and large radii. This is partly due to the symmetries ($\bar J_V^r$, for instance, has to vanish for $r \to 0$) and the condition $u_\mu \bar J_V^\mu=0$. But here the situation is more complicated due to the interplay of time evolution and the spacial matter distribution. For instance, jets traveling outwards encounter less material than their partners going inwards, i.e.\ towards the center of the overlap region, which contributes to $\bar J_V^r$ at intermediate and late times.

\begin{figure}[ht]
\hspace*{-1cm}
\includegraphics[angle=-90,width=0.55\linewidth]{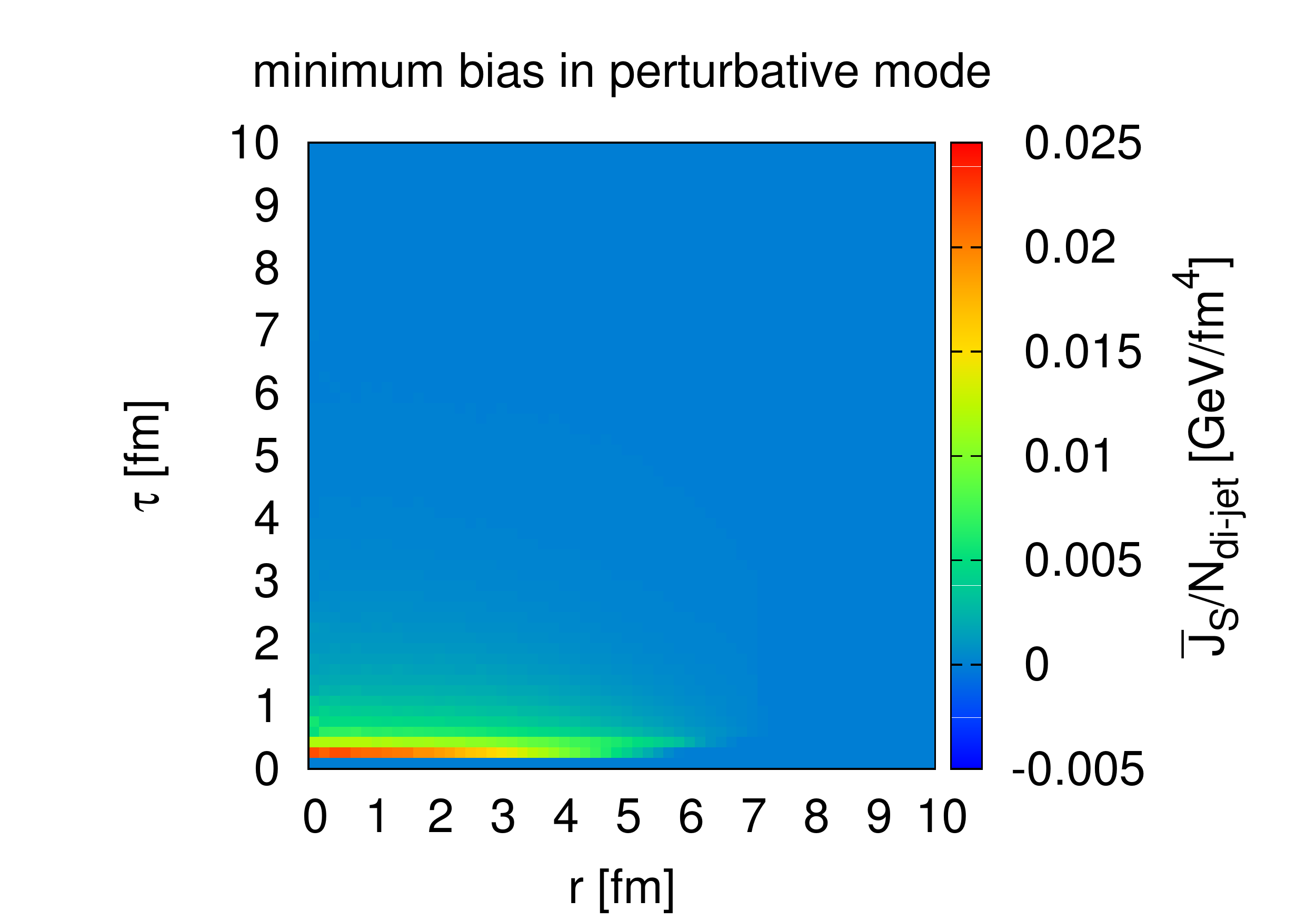}
\hspace*{-1cm}
\includegraphics[angle=-90,width=0.55\linewidth]{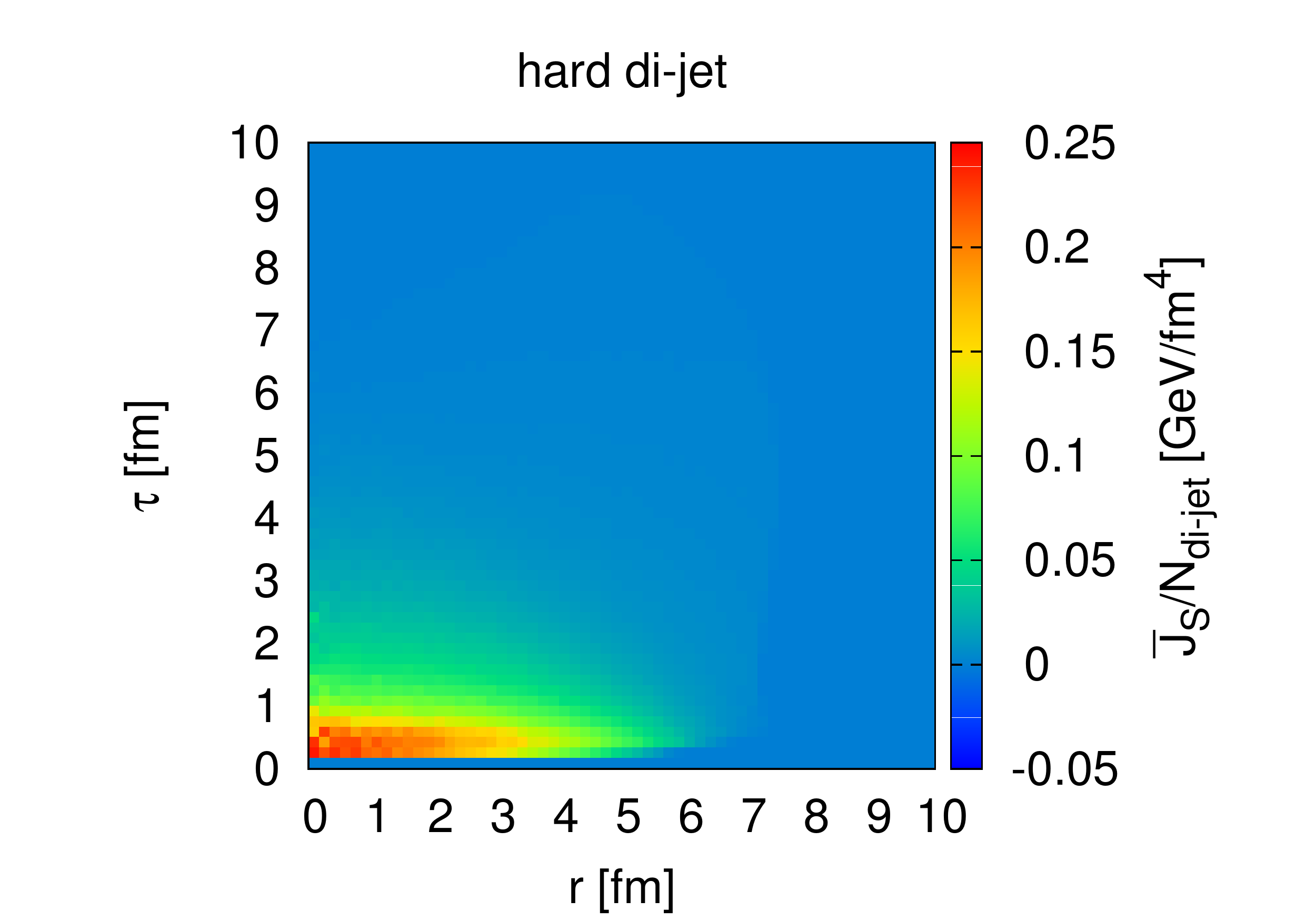}\par
\hspace*{-1cm}
\includegraphics[angle=-90,width=0.55\linewidth]{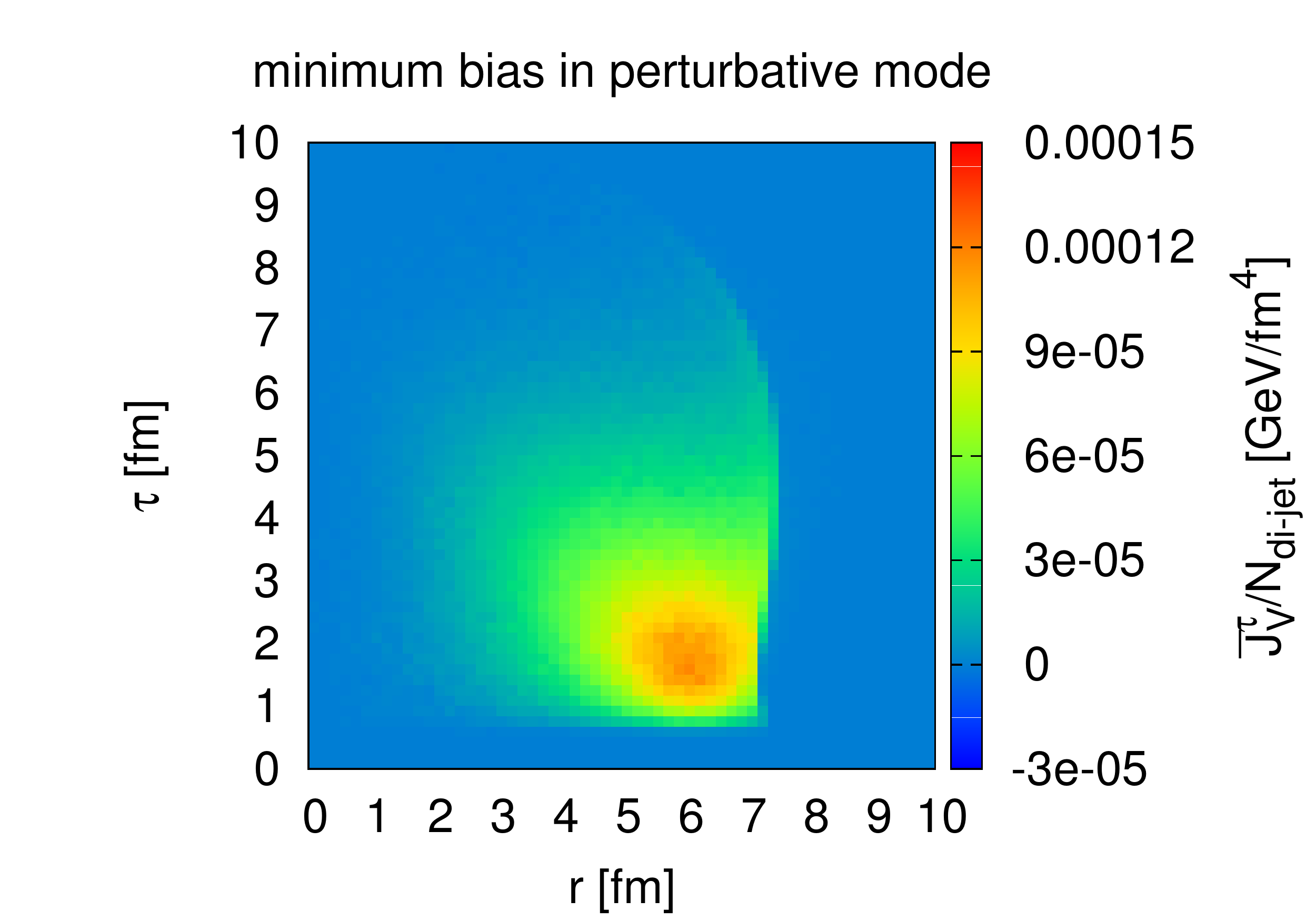}
\hspace*{-1cm}
\includegraphics[angle=-90,width=0.55\linewidth]{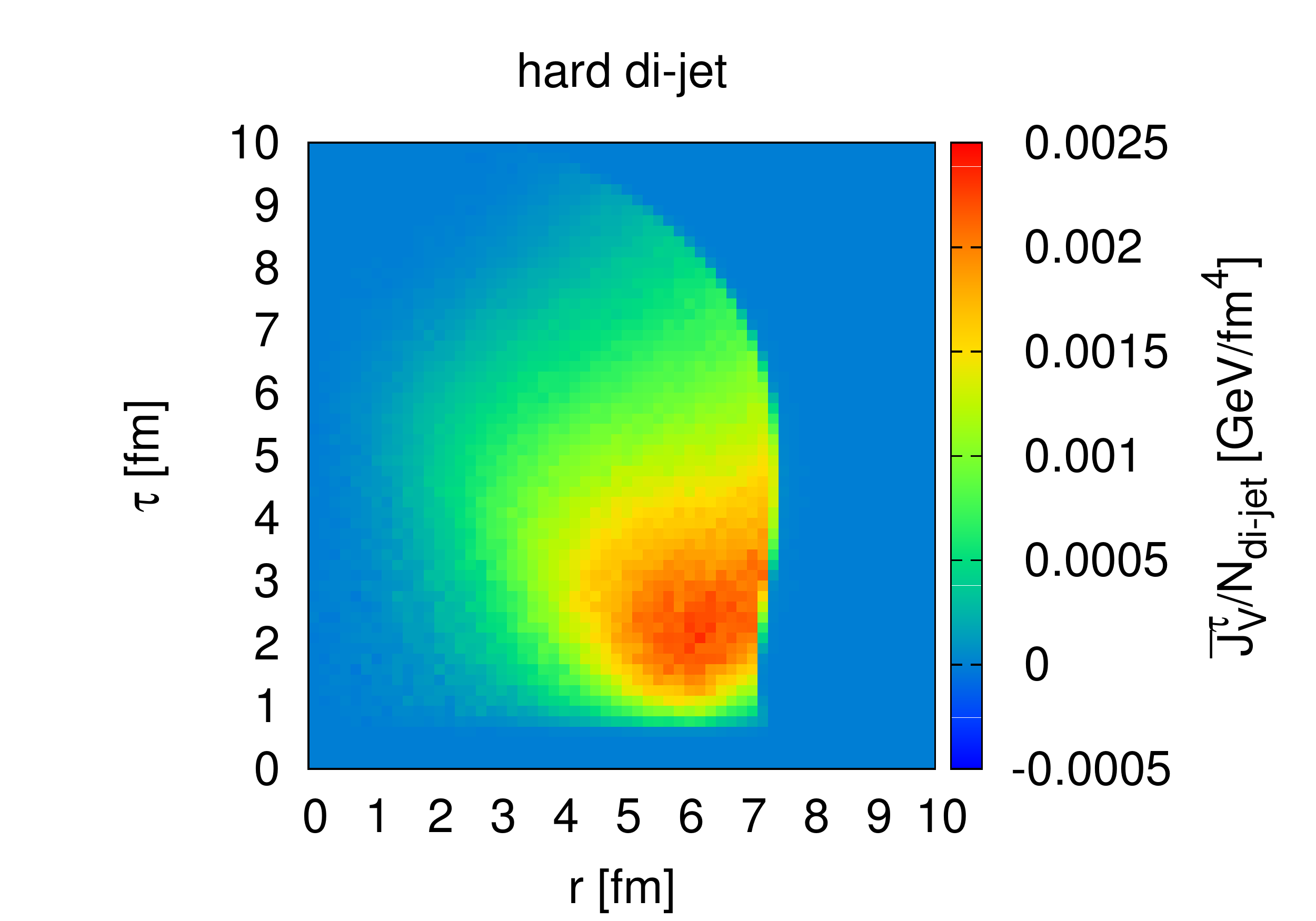}\par
\hspace*{-1cm}
\includegraphics[angle=-90,width=0.55\linewidth]{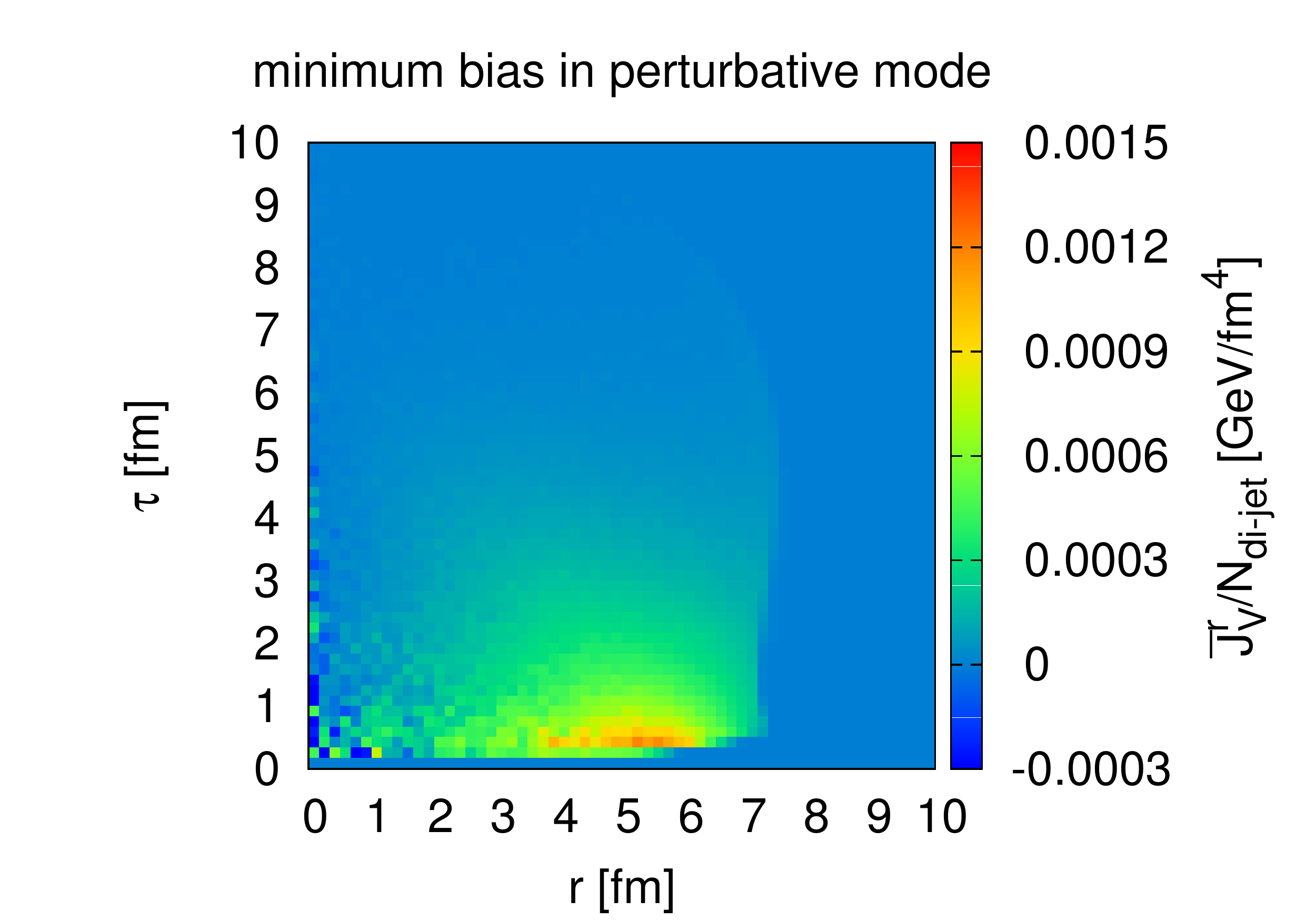}
\hspace*{-1cm}
\includegraphics[angle=-90,width=0.55\linewidth]{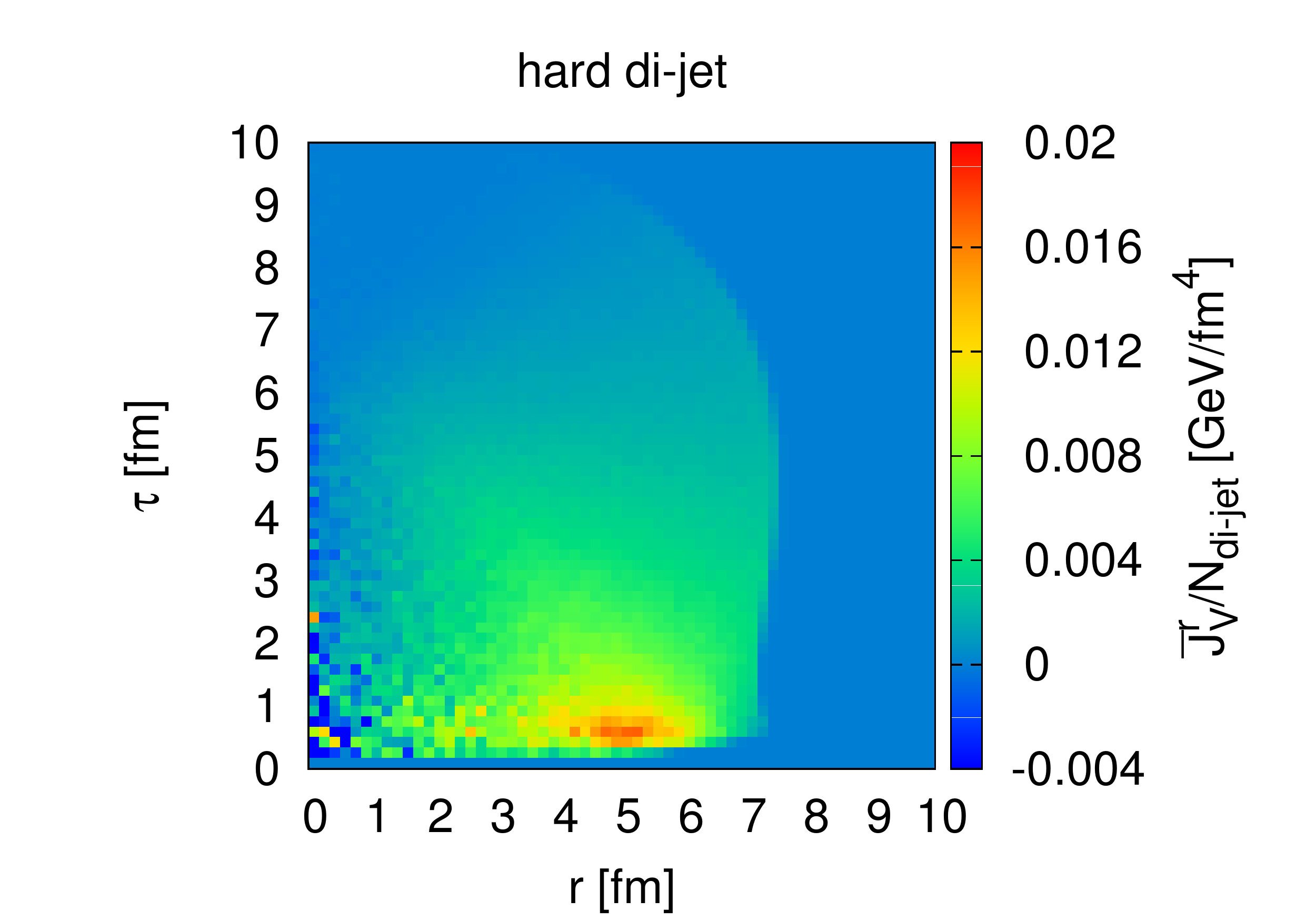}\par
\caption{$\bar J_S$ (top row), $\bar J_V^\tau$ (middle row) and $\bar J_V^r$ (bottom row) per di-jet for the minimum bias scenario in the perurbative mode (left column) and the hard di-jet scenario (right column).}
\label{fig::j100s}
\end{figure}

\begin{figure}
\hspace*{-1cm}
\includegraphics[angle=-90,width=0.55\linewidth]{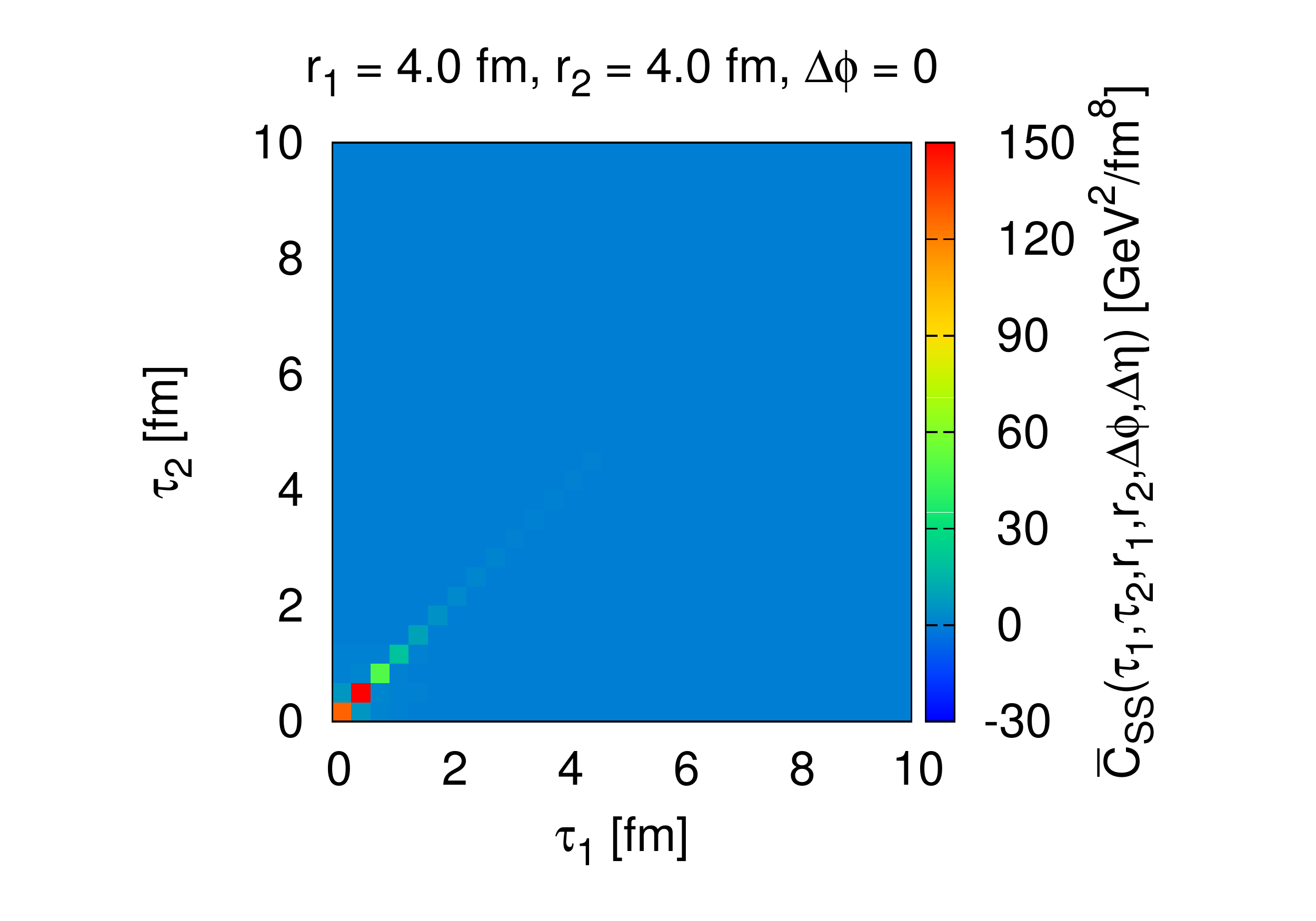}
\hspace*{-1cm}
\includegraphics[angle=-90,width=0.55\linewidth]{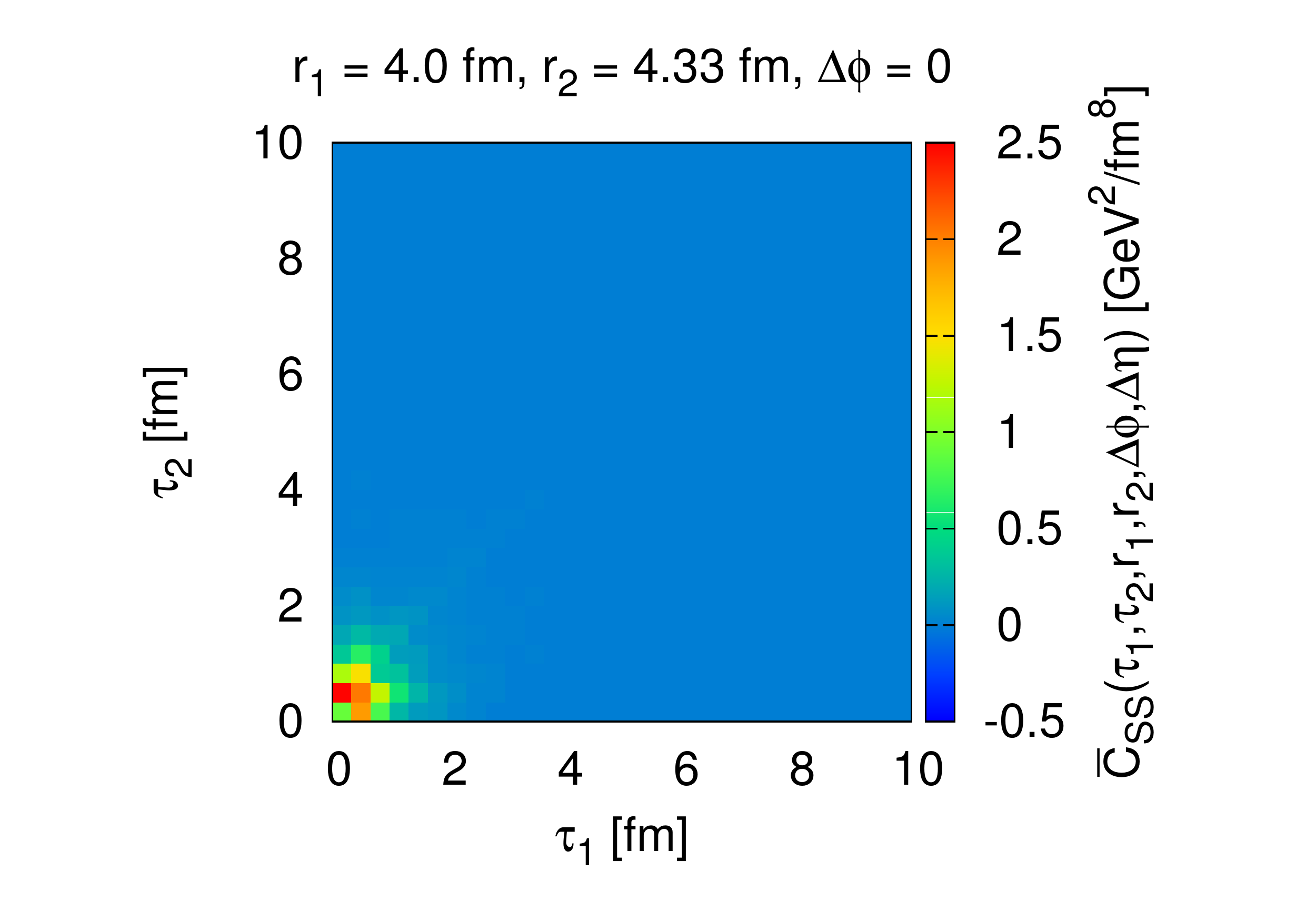}\par
\hspace*{-1cm}
\includegraphics[angle=-90,width=0.55\linewidth]{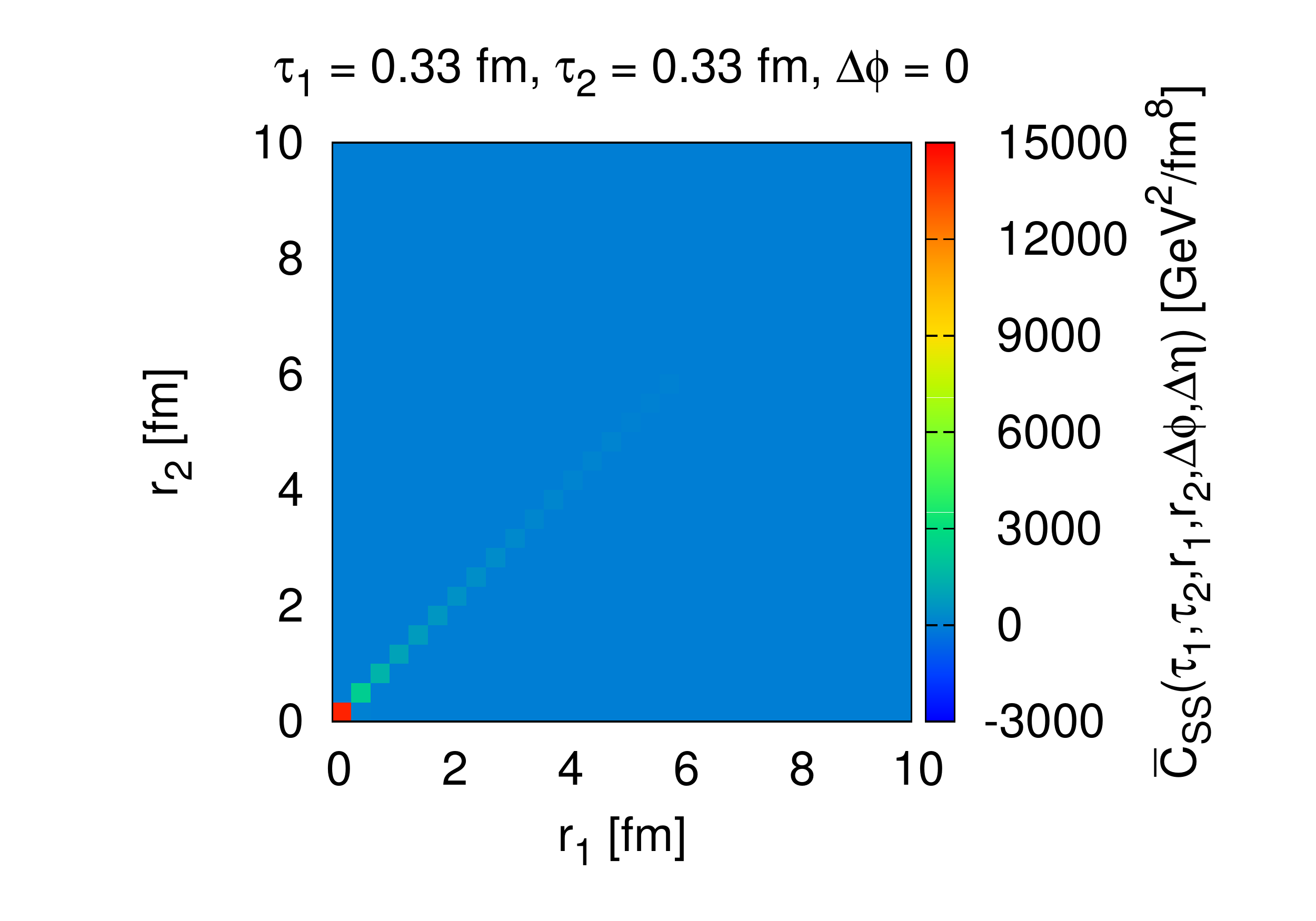}
\hspace*{-1cm}
\includegraphics[angle=-90,width=0.55\linewidth]{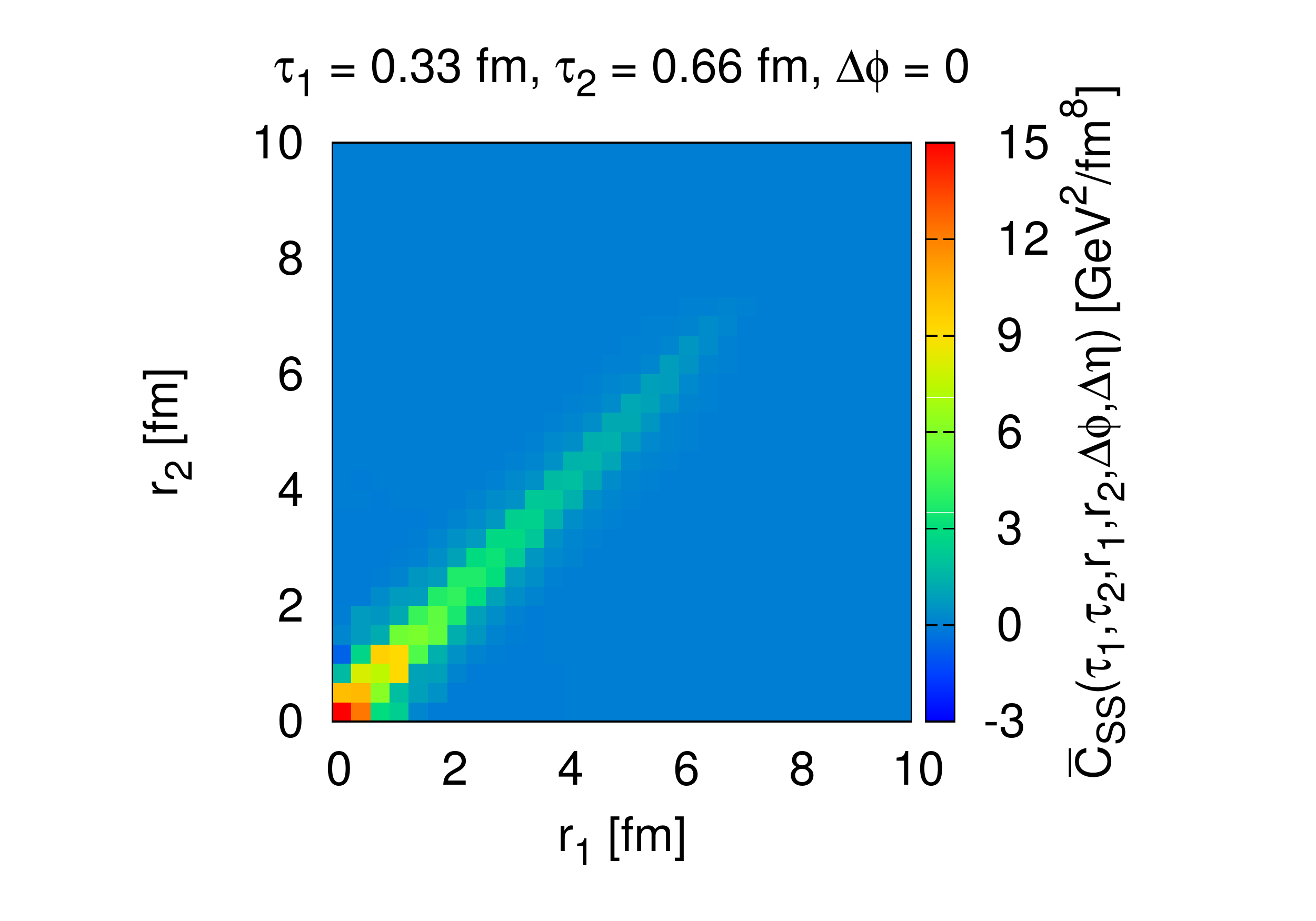}\par
\includegraphics[angle=-90,width=0.49\linewidth]{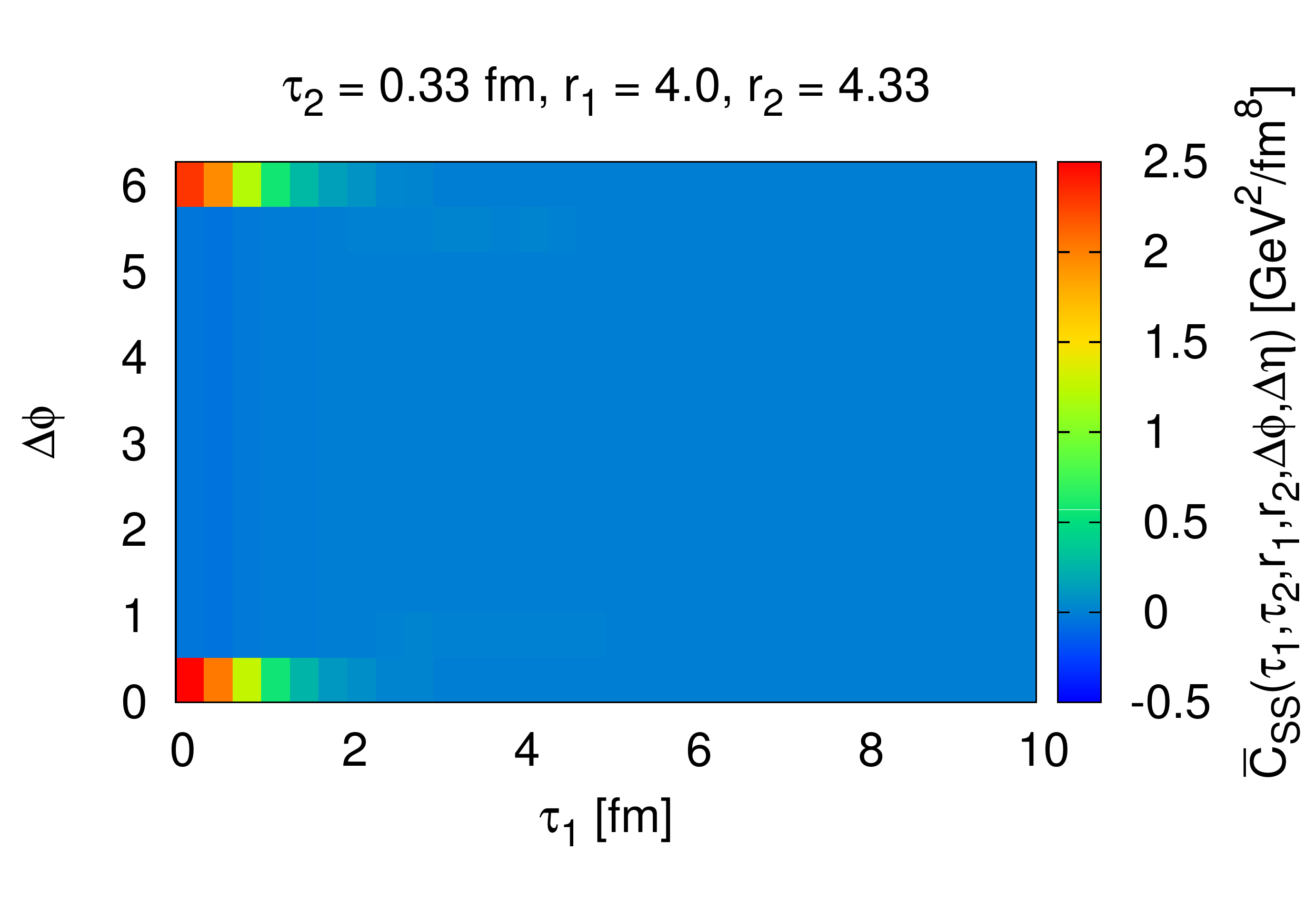}
\includegraphics[angle=-90,width=0.49\linewidth]{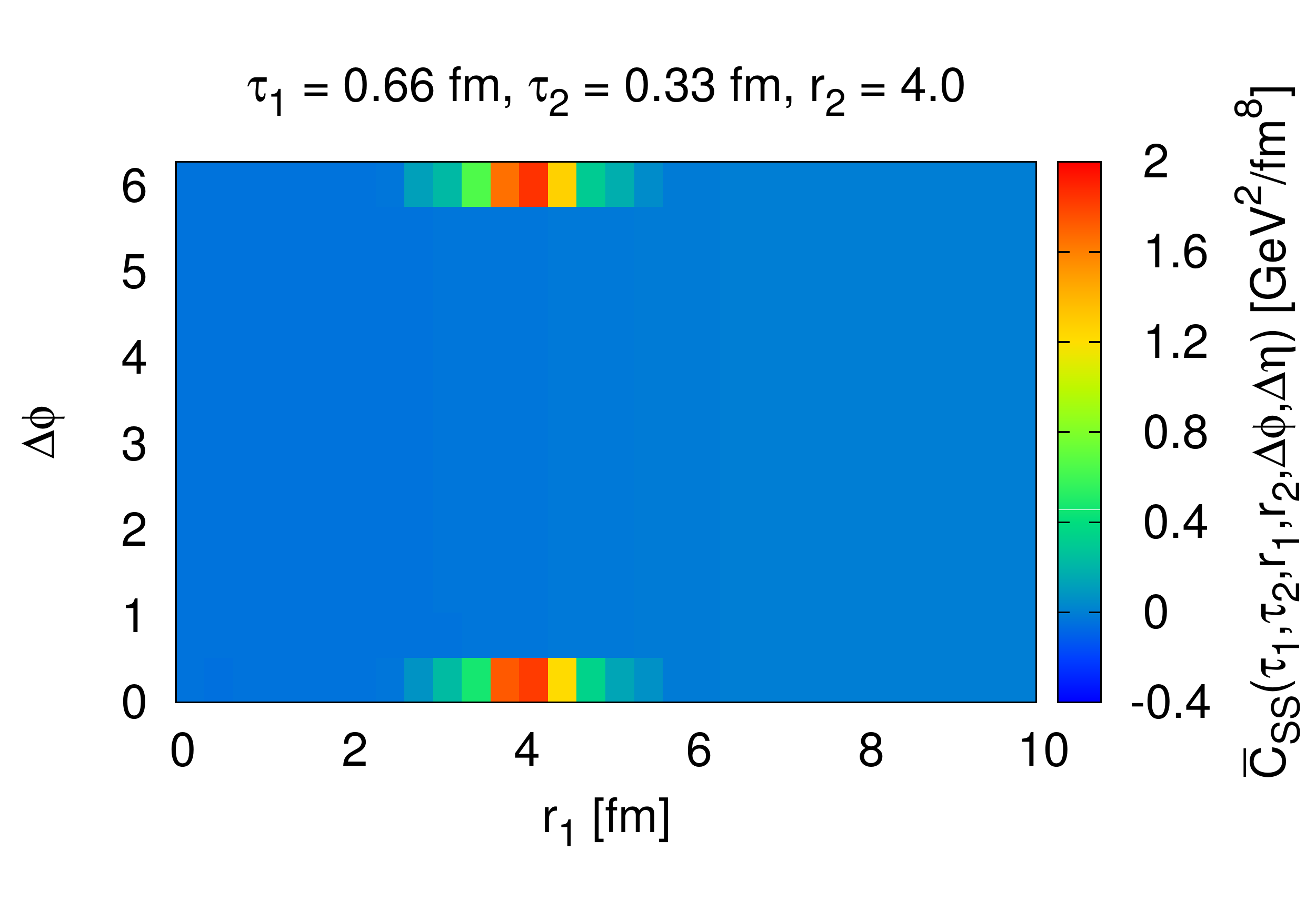}\par
\caption{Scalar correlation function $\bar C_{SS}$ for the perturbative minimum bias scenario. The $\Delta \eta$ dependence is averaged over.}
\label{fig::css}
\end{figure}

\begin{figure}
\hspace*{-1cm}
\includegraphics[angle=-90,width=0.55\linewidth]{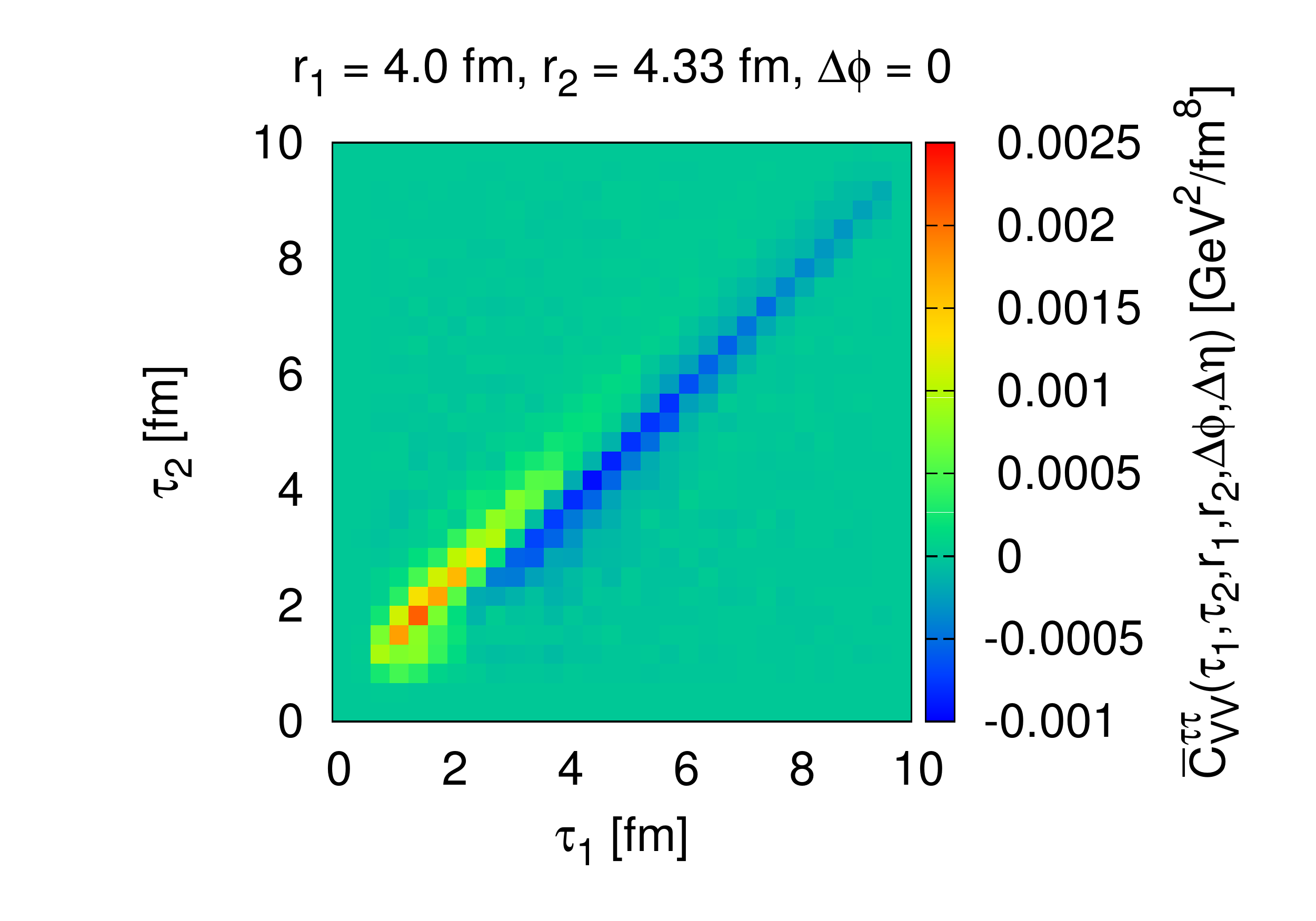}
\hspace*{-1cm}
\includegraphics[angle=-90,width=0.55\linewidth]{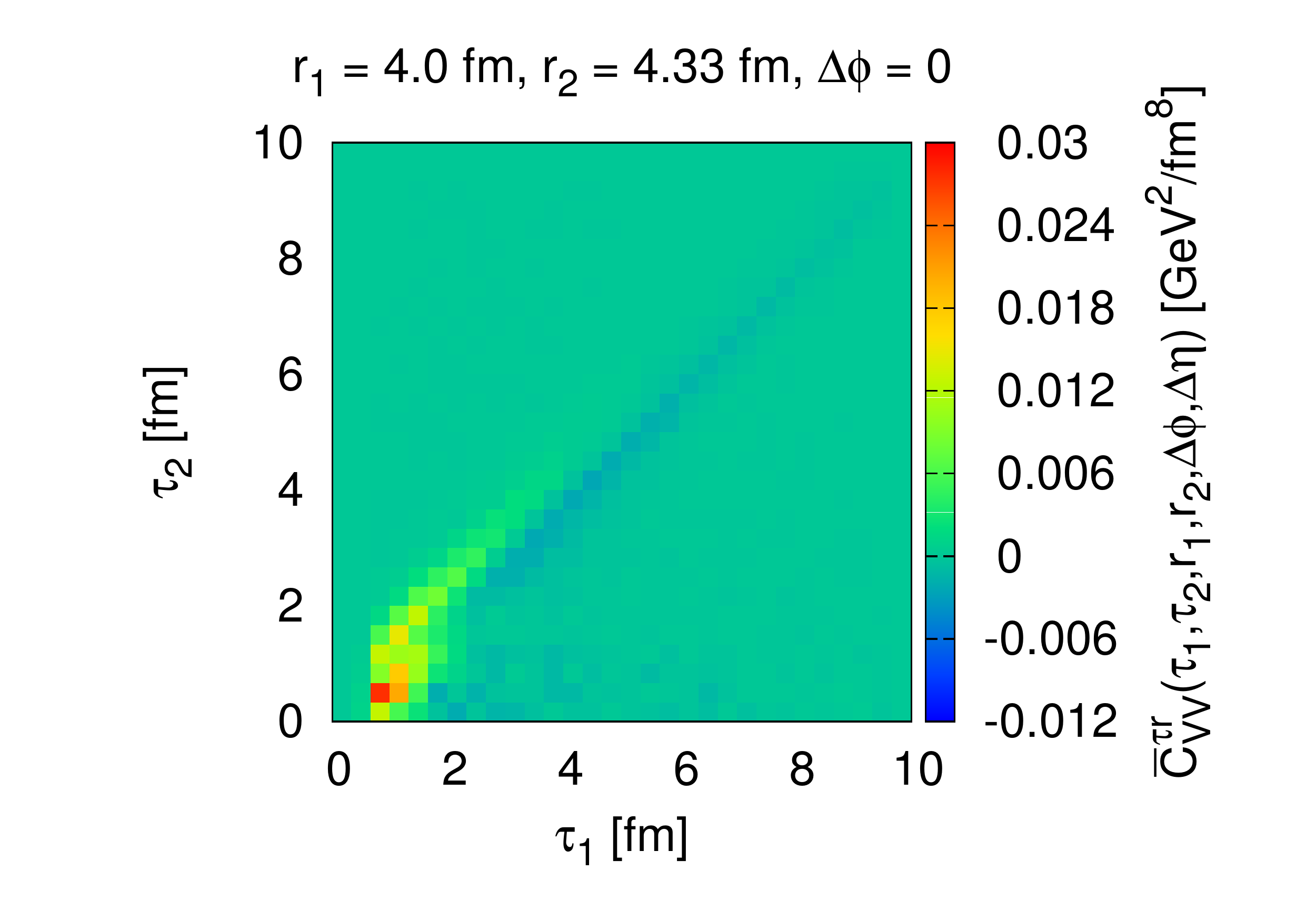}\par
\caption{$\bar C_{VV}^{\tau\tau}$ and $\bar C_{VV}^{\tau r}$ component of the tensor correlator for the perturbative minimum bias scenario. The $\Delta \eta$ dependence is averaged over.}
\label{fig::cvv}
\end{figure}

Figures~\ref{fig::2js} to \ref{fig::cvv} also confirm the expectation that the source term is on average positive also in the minimum bias scenarios, i.e.\ energy and momentum flow from the jets to the background. Figures~\ref{fig::2js} and \ref{fig::jv} are based on a sample of $8\cdot 10^6$ Monte-Carlo di-jets and the oscillating features that are visible in some of the curves are numerical fluctuations.

\medskip

In the hard di-jet scenario a cut of $p_{\perp,\text{cut}} = \unit[100]{GeV}$ is placed on the matrix element. The final jet population looks very different due to quenching of the jets. When comparing to experimental data one would have to place the cut on the final jet $\pt$, which is straightforward but not necessary for this exploratory study. In figure~\ref{fig::j100s} the momentum deposition of such a hard di-jet is compared to a minimum bias di-jet. As expected, the source term of hard jets is much larger in magnitude and extends to significantly later times. The energy transfer $\bar J_S$ from hard di-jets follows the approximate scaling with $T^2(\tau,r)\cdot N_\text{coll}(r)$ for much longer since they don't reach thermal scales quickly. Deviations come from the dilution of the $N_\text{coll}(r)$ distribution due to the propagation of the jets and an increase in the multiplicity of hard partons due to splitting. The same effects are also at work in $\bar J_V$.

To obtain the source term for the entire event one has to add to the contribution of the hard di-jet $N_\text{di-jet}-1$ times that of a minimum bias jet. A  $\mathcal{O}(\unit[100]{GeV})$ di-jet deposits roughly a factor of 40 more  energy and momentum than a minimum bias jet. Since the number of minimum bias di-jets per event is of the order 1500, the presence of a hard di-jet increases the energy transfer per event only by about $\unit[2-3]{\%}$.

The influence of the expectation values $\bar J_S$ and $\bar J_V$ in the perturbative minimum bias scenario onto the fluid dynamic evolution are discussed in section \ref{sec:TheHydrodynamicEvolution}, see in particular figs.\ \ref{fig::temp} and \ref{fig::fluidVelocity}. As the effect is already small in this case, the contribution of an additional hard di-jet is negligible for all practical purposes.

\bigskip

For the correlation functions of source terms as defined in eqs.\ \eqref{eq:corrfunctsource} it is convenient to take the event average and subtract the disconnected contribution,
\begin{align}
\bar C_{SS}(x,y) & = \mean{J_S(x)J_S(y)} - J_S(x) J_S(y) \\
\bar C_{SV}^\mu(x,y) & = \mean{J_S(x)J_V^\mu(y)} - J_S(x) J_V^\mu(y) \\
\bar C_{VV}^{\mu\nu}(x,y) & = \mean{J_V^\mu(x)J_V^\nu(y)} - J_V^\mu(x) J_V^\nu(y) \,.
\label{eq:conncorrfunct}
\end{align}
In this form the correlators also scale linearly with $N_\text{di-jet}$. They depend on $\tau_1$, $\tau_2$, $r_1$, $r_2$, $\Delta \phi$ and $\Delta \eta = |\eta_1 - \eta_2|$. We again average over $\Delta \eta$ in the rapidity window under consideration, but keep the dependence on $\Delta \phi$. The scalar correlator $\bar C_{SS}$ is shown in figure~\ref{fig::css} for the perturbative minimum bias scenario. It is positive everywhere (except for statistical fluctuations) and strongly peaked at $x=y$. The fact that the correlation functions in eq.\ \eqref{eq:conncorrfunct} decay quickly with the separation between the arguments $x-y$ implies that the corresponding fluctuations that influence the fluid dynamics are essentially local. Similarly to the averaged source terms, the correlation functions are largest at small times $\tau$ and radii $r$. They extend to radii of about $\unit[6]{fm}$ whereas the decay with time is very fast. The minimum-bias and hard di-jet samples look similar, for the latter the correlation functions are slightly broader. Figure~\ref{fig::cvv} shows the $\tau\tau$ and $\tau r$ component as examples of the tensor $\bar C_{VV}^{\mu\nu}$. Here positive as well as negative structures develop.

The correlators generate potentially sizeable contributions to correlation observables on the fluid dynamic side such as the anisotropic flow coefficients $v_n$. The calculations are somewhat more involved than for the averaged source terms and will be discussed in a separate publication.

\section{Conclusions}
\label{sec:Conclusions}

We have studied the influence of the energy deposition by jets onto the evolution of the medium by combing a realistic microscopic jet quenching model with a fluid dynamic description of the bulk. The energy-momentum transfer from jets constitutes source terms in the hydrodynamic evolution equations, which we characterize in terms of event averages and correlation functions. 
The event averaged source function that enters the time evolution of energy density is largest at early times and for small radii. It leads to an increase of temperature due to the additional dissipated energy but the effect is numerically very small.
The expectation value for the source function that enters the evolution of the radial fluid velocity peaks at intermediate times (a few fm/c) and for large radii (about \unit[6]{fm/c}). This term gives the effect of the force opposing drag and leads to an increase of radial flow of up to about 10\%: The jets drag the fluid outwards.
The momentum transfer causing the increase in radial flow was shown to have a non-trivial functional form, which is not easily captured by simple parametrisations of the source term. This highlights the advantage of constructing a realistic source term using a model based on microscopic dynamics.

Our formalism allows to study also event-by-event fluctuations in the source terms. Here we quantify connected two-point correlation functions and find that they are largest at early times and that they are local (the correlation functions peak strongly for equal space-time arguments). 

A conceptual difficulty of a formalism that combines a microscopic description of jets with a macroscopic description of the medium is that the separation between the two components is to some extent arbitrary. This becomes apparent in the difficulties related to defining the jet population. We chose to regularise the perturbative jet cross section such that jets are produced predominantly in the phase space region where they dominate over the thermal distribution. This leads to a rather low $\pt$ cut-off of the order a few GeV. In this region the perturbative cross section has large uncertainties and multi-parton interactions may play a role. These difficulties are extenuated to some extent by the fact that -- due to the fact that we employ a dynamical  model of jet quenching -- very soft partons on average do not lose energy and thus do not contribute to the fluid dynamic source terms. Nevertheless, we estimate the resulting uncertainties for the latter to be of the order of a factor 2 or 3.

We studied the source terms generated in a 'typical' event, i.e.\ without cuts on the jets, and the effect of a high-$\pt$ ($\mathcal{O}(\unit[100]{GeV})$) di-jet. In events containing a hard di-jet the energy and momentum deposition is increased by only a few percent as compared to the minimum bias scenario. The presence of a hard jet is thus negligible for global observables. This can be different for correlation observables such as harmonic flow coefficients, which receive potentially sizable contributions from jets. These will be studied in an upcoming publication.

\section*{Acknowledgements}
We would like to thank J.~G.~Milhano and U.~A.~Wiedemann for valuable comments on the manuscript.

\end{document}